\documentclass[12pt,preprint]{aastex}

\newcommand{\acsb}{$B_{\rm 435}$}
\newcommand{\acsv}{$V_{\rm 606}$}
\newcommand{\acsi}{$i_{\rm 775}$}
\newcommand{\acsz}{$z_{\rm 850}$}
\newcommand{\bviz}{$B_{\rm 435}, V_{\rm 606}, i_{\rm 775}, z_{\rm 850}$}
\newcommand{\ks}{$K_{\rm s}$}
\newcommand{\jm}{$J$}
\newcommand{\hm}{$H$}
\newcommand{\jw}{$J$}
\newcommand{\wcam}{WIRCam}
\newcommand{\moircs}{MOIRCS}
\newcommand{\zdrop}{$z\gtrsim 6.5$}
\newcommand{\sqarc}{arcmin$^2$}

\newcommand{\etal}{{et\thinspace al.}}
\newcommand{\sn}{$S/N$}
\newcommand{\Ho}{$H_{0}$}

\newcommand{\tabref}[1]{Table~\ref{#1}}
\newcommand{\figref}[1]{Figure~\ref{#1}}
\newcommand{\secref}[1]{\S~\ref{#1}}

\begin{document}

\title{Near-Infrared Survey of the GOODS-North Field: Search for
  Luminous Galaxy Candidates at \zdrop\ \altaffilmark{\ast\dag}
}

\shorttitle{Near-IR imaging in GOODS-N}

\author{
Nimish~P.~Hathi\altaffilmark{1,2}, 
Bahram~Mobasher\altaffilmark{2}, 
Peter~Capak\altaffilmark{3},
Wei-Hao~Wang\altaffilmark{4},
Henry~C.~Ferguson\altaffilmark{5}
}

\altaffiltext{1}{Observatories of the Carnegie Institution for
  Science, Pasadena, CA 91101, USA}

\altaffiltext{2}{Department of Physics and Astronomy, University of
  California, Riverside, CA 92521, USA}

\altaffiltext{3}{Department of Astronomy, 249-17 Caltech, 1201
  E. California Blvd., Pasadena, CA 91125, USA}

\altaffiltext{4}{Institute of Astronomy and Astrophysics, Academia
  Sinica, P.O. Box 23-141, Taipei 10617, Taiwan}

\altaffiltext{5}{Space Telescope Science Institute, 3700 San Martin
  Drive, Baltimore, MD 21218, USA}

\altaffiltext{$\ast$}{Based in part on
    data collected at Subaru Telescope, which is operated by the
    National Astronomical Observatory of Japan.}

\altaffiltext{$\dag$}{Based on
    observations obtained with WIRCam, a joint project of CFHT,
    Taiwan, Korea, Canada, France, at the Canada-France-Hawaii
    Telescope (CFHT) which is operated by the National Research
    Council (NRC) of Canada, the Institute National des Sciences de
    l'Univers of the Centre National de la Recherche Scientifique of
    France, and the University of Hawaii.}

\email{nhathi@obs.carnegiescience.edu}
\shortauthors{Hathi et al}

\begin{abstract}
 
  We present near-infrared (NIR; $J$ \& \ks) survey of the Great
  Observatories Origins Deep Survey-North (GOODS-N) field. The
  publicly available imaging data were obtained using the MOIRCS
  instrument on the 8.2m Subaru and the WIRCam instrument on the 3.6m
  Canada-France-Hawaii Telescope (CFHT).  These observations fulfill a
  serious wavelength gap in the GOODS-N data - i.e., lack of deep NIR
  observations. We combine the Subaru/MOIRCS and CFHT/WIRCam archival
  data to generate deep $J$ and \ks\ band images, covering the full
  GOODS-N field ($\sim$169~\sqarc) to an AB magnitude limit of
  $\sim$25~mag (3$\sigma$).  We applied \acsz-band dropout color
  selection criteria, using the NIR data generated here. We have
  identified two possible Lyman Break Galaxy (LBG) candidates at
  \zdrop\ with $J \lesssim 24.5$. The first candidate is a likely LBG
  at $z\simeq 6.5$ based on a weak spectral feature tentatively
  identified as Ly$\alpha$ line in the deep Keck/DEIMOS spectrum,
  while the second candidate is a possible LBG at $z\simeq 7$ based on
  its photometric redshift.  These \acsz-dropout objects, if
  confirmed, are among the brightest such candidates found so far. At
  \zdrop, their star formation rate is estimated as
  100--200~M$_\odot$~yr$^{-1}$. If they continue to form stars at this
  rate, they assemble a stellar mass of
  $\sim$5$\times$10$^{10}$~M$_\odot$ after about 400 million years,
  becoming the progenitors of massive galaxies observed at $z\simeq
  5$. We study the implication of the \acsz-band dropout candidates
  discovered here, in constraining the bright-end of the luminosity
  function and understanding the nature of high redshift galaxies.

\end{abstract}



\section{Introduction}\label{introduction}

In recent times, the high redshift frontier has started to probe the
epoch of reionization because of extensive observations at
Near-Infrared (NIR) wavelengths.  This extension of multi-wavelength
galaxy surveys to obtain deep images at observed NIR has provided a
new prospect for the study of galaxies at high ($z\gtrsim 6.5$)
redshifts. Such observations are vital to identify star-forming high
redshift objects by sampling their rest-frame UV light, and for
estimating their star formation properties. On the other hand, the
combination of \emph{Spitzer}/IRAC and NIR observations can be used to
identify massive, evolved Balmer Break Galaxies (BBGs) at $z\gtrsim 5$
\citep[e.g.,][]{wikl08}. Furthermore, relative insensitivity at these
wavelengths to the stellar population mix, dust and K-correction,
specially at intermediate ($z\simeq 2$) redshifts is an added bonus.

Among the most extensive datasets available in extragalactic
astronomy, is the Great Observatories Origins Deep Survey
\citep[GOODS;][]{giav04a}. This consists of imaging of two fields
(GOODS-North and GOODS-South), by the \emph{Hubble Space Telescope}
(\emph{HST}) Advanced Camera for Surveys (ACS) in four optical bands
(\acsb, \acsv, \acsi, \acsz). Given the need for NIR data in any study
of evolution of galaxies, the GOODS-South (GOODS-S) field has been
observed extensively at NIR wavelengths, using the 8.2m Very Large
Telescope \citep[VLT;][]{renz03} in the $J$, $H$ and \ks\ bands, but
the GOODS-North (GOODS-N) field has not had such full-field deep
coverage at these wavelengths. In the last few years, the Wide-field
InfraRed Camera \citep[\wcam;][]{puge04} on the 3.6m
Canada-France-Hawaii Telescope (CFHT), and the Multi-Object Infrared
Camera and Spectrograph \citep[\moircs;][]{ichi06,suzu08} on the 8.2m
Subaru telescope have invested large amounts of observing time to
obtain NIR observations of the GOODS-N field.  The \wcam\ \ks\ data
were published by \citet{wang10} and the \moircs\ $J$, $H$, and \ks\
data were published by \citet{kaji11}.  In both works, the images and
catalogs were made publicly available.  Various studies have used
subsets of these observations for specific science goals.  For
example, \citet{kaji06} used \moircs\ data to investigate the number
counts of Distant Red Galaxies, while \citet{ichi07} used \moircs\
\ks-band selected galaxies to measure clustering properties of
low-mass galaxies at $1< z < 4$. \citet{bund09} used NIR observations
from both GOODS fields to investigate the dependence on stellar mass
and galaxy type of the close pair fractions and implied merger rate,
while \citet{bouw08} used the deepest NIR observations from both
fields to search for star-forming galaxies at $z\gtrsim 6.5$ and
constrain their rest-frame ultraviolet (UV) luminosity functions.
These studies show the significance of NIR observations in such
well-studied fields. The full coverage of the GOODS-N region is
essential not only to expand these studies, but also to accomplish
larger number of additional science goals. Despite such studies, there
is limited information in literature of the observational details and
sensitivities of the NIR data for GOODS-N field, and details about
source detection.

The high sensitivity of the \emph{HST}/Wide Field Camera 3 (WFC3) has
made it possible to identify faint Lyman break galaxies (LBGs) at
$z\gtrsim 6.5$
\citep[e.g.,][]{oesc10,bouw10a,bouw10b,fink10,yan10,wilk11}, when the
universe was only about 500 Myr old. However, because most of these
candidates are very faint (AB$\,\gtrsim\,$27~mag), it is difficult to
measure their spectra or measure their rest-frame optical SEDs to
better constrain their stellar population and mass.  It is imperative
to search for relatively brighter (AB$\,\lesssim\,$25~mag) galaxies at
$z\gtrsim 6.5$, and study them in great detail. There are three
principal ways to accomplish this goal. First approach is through big
cluster surveys, which are searching for lensed galaxies at
\zdrop\ \citep[e.g.,][]{hall12,brad12}. These galaxies are
intrinsically faint but because of the magnification their observed
magnitudes are relatively bright.  Second approach is to take
advantage of the \emph{HST} parallel observations
\citep[e.g,.][]{tren11,yan11} to explore large number of random fields
at relatively shallow depth, and the final approach is to perform wide
area ground-based surveys \citep[e.g.,][]{ouch09, cast10,
  hick10,capa11} to identify and study such bright candidates.  These
methods are starting to identify bright galaxy candidates at \zdrop\
and are putting important constraints on the bright end of the
rest-frame UV luminosity functions, though
the number of these candidates remains too small to make any
statistically significant conclusions about their physical properties.

The wide area ground-based surveys have few distinct advantages
compared to space-based approaches of identifying bright high redshift
(\zdrop) galaxies. First, they can cover much larger area than
space-based observations, and secondly, the \ks-band data, which is
important for accessing the reliability of high redshift candidates,
is only available through ground-based telescopes. These reasons make
NIR ground-based surveys very appealing for searching bright high
redshift galaxies.

Here, we present NIR data in the GOODS-N field obtained using the
CFHT/\wcam\ and the Subaru/\moircs\ instruments. We combine the
archival data from these instruments to generate deeper images in $J$-
and \ks-bands. The Subaru/\moircs\ data cover $\sim$60\% of the
central GOODS-N region in $J$, $H$, \ks-filters, while the CFHT/\wcam\
covers a much larger area, extending well beyond the GOODS-N field in
the $J$, \ks-filters. The combination of the \moircs\ and \wcam\ NIR
data covers full area of the GOODS-N field ($\sim$169~\sqarc), and is
deeper in the central $\sim$60\% of the field.  These combined images
will extend far beyond the traditional field which the \emph{HST}/WFC3 data
will cover, and will considerably help to undertake multi-wavelength
science goals that require NIR observations. As one application, here
we use these NIR images along with the \emph{HST}/ACS \bviz\ data to identify
bright \acsz-dropouts (i.e., LBG candidates at \zdrop), and estimate
their surface density in the GOODS-N field at the brighter limits
(AB$\,<\,$25~mag).

We describe both the \moircs\ and \wcam\ NIR observations along with
existing \emph{HST}/ACS data in \secref{obs}, the NIR data reduction
is discussed in \secref{data}, and the quality of our reduced NIR data
is showed in \secref{quality}. We also discuss the \ks-selected
catalogs, their number counts and the completeness estimates in
\secref{quality}. In \secref{lbgs}, we develop and implement
\acsz-dropout selection criteria to identify LBG candidates at
\zdrop\ and discuss the implications of our results. We summarize our
results in \secref{summary}, emphasizing that we combine two NIR
surveys of GOODS-N field to construct deep \ks-band images for this
field, needed for a variety of studies.

Throughout this paper we refer to the \emph{HST}/ACS F435W, F606W,
F775W, F850LP filters as \acsb, \acsv, \acsi, \acsz, and to the
\emph{Spitzer}/IRAC 3.6~$\mu$m, 4.5~$\mu$m, 5.8~$\mu$m, 8.0~$\mu$m
filters as [3.6], [4.5], [5.8], [8.0], respectively, for convenience.
We assume a \emph{Wilkinson Microwave Anisotropy Probe} (WMAP)
cosmology with $\Omega_m$=0.274, $\Omega_{\Lambda}$=0.726 and
\Ho=70.5~km~s$^{-1}$~Mpc$^{-1}$, in accord with the 5 year WMAP
estimates of \citet{koma09}. This corresponds to a look-back time of
12.93~Gyr at $z \simeq 7$.  Magnitudes are given in the AB system
\citep{oke83}.


\section{Observations}\label{obs}
\subsection{Subaru/\moircs\ Observations}

\begin{figure}
\epsscale{1.0}
\plotone{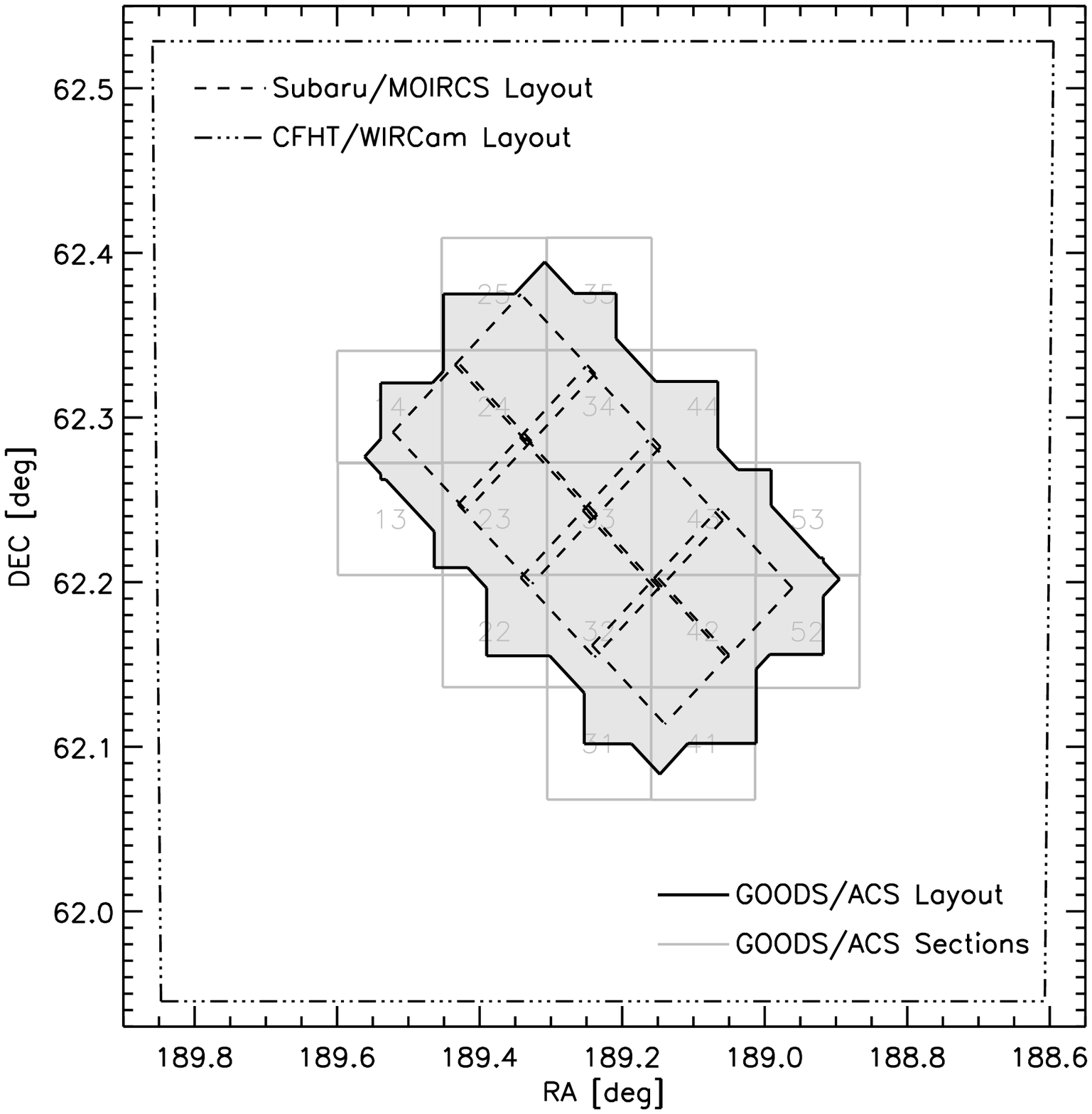}
\caption{The Subaru/\moircs\ (dashed) and the CFHT/\wcam\ (large box
  shown with dot-dash) field layouts in comparison to the
  GOODS-N ACS (solid) field. The GOODS-N ACS field (a large grid of
  40K$\times$40K pixels) was divided into 17 sections (each
  8K$\times$8K pixels) for data handling convenience. The gray
  solid lines show layout of these sections.}\label{fig:layout}
\end{figure}

The \moircs\ is on the Subaru 8.2m telescope, providing wide-field
imaging and spectroscopic capability in the NIR bands.  In the imaging
mode, the \moircs\ provides 4$\times$7 \sqarc\ field of view with
0.117\arcsec\ pixel$^{-1}$ using two 2048$\times$2048 HgCdTe (HAWAII2)
arrays, each of which covers a 4$\times$3.5 \sqarc\ field. There is
almost no gap between FOVs of each channel. Details of this instrument
are given in \citet{ichi06}.  We used the Subaru archive
SMOKA\footnote{http://smoka.nao.ac.jp/} to retrieve the \moircs\
observations in the GOODS-N field. Typical exposure time for
individual frames varied between $\sim$50 to $\sim$100 secs and
standard nine-point dither patterns (dithering by 10-15\arcsec) were
used \citep[for details see, e.g.,][]{bund09,kaji11}. These
observations were carried out over a period of three years between
2005--2007 \citep[e.g.,][]{wang09,bund09,kaji11}. They cover an area
of about 109~\sqarc, which overlaps with $\sim$60\% of the central
region of the GOODS-N ACS field, as shown in \figref{fig:layout}. The
typical seeing for these observations is between 0.6\arcsec\ and
0.8\arcsec. \figref{fig:layout} shows the GOODS-N ACS field in gray
shaded region with black solid line border, and the \moircs\ covered
field with dashed black line. The full \moircs\ field is covered by 4
pointings and the field layout is shown by 2$\times$4 chips
(\figref{fig:layout}).  These observations were done in three
broad-band filters \jm, \hm\ and \ks, whose transmission curves are
shown as solid curves in the top panel of the \figref{fig:filters}.

\subsection{CFHT/\wcam\ Observations}

\begin{figure}
\begin{center}
\epsscale{0.725}
\plotone{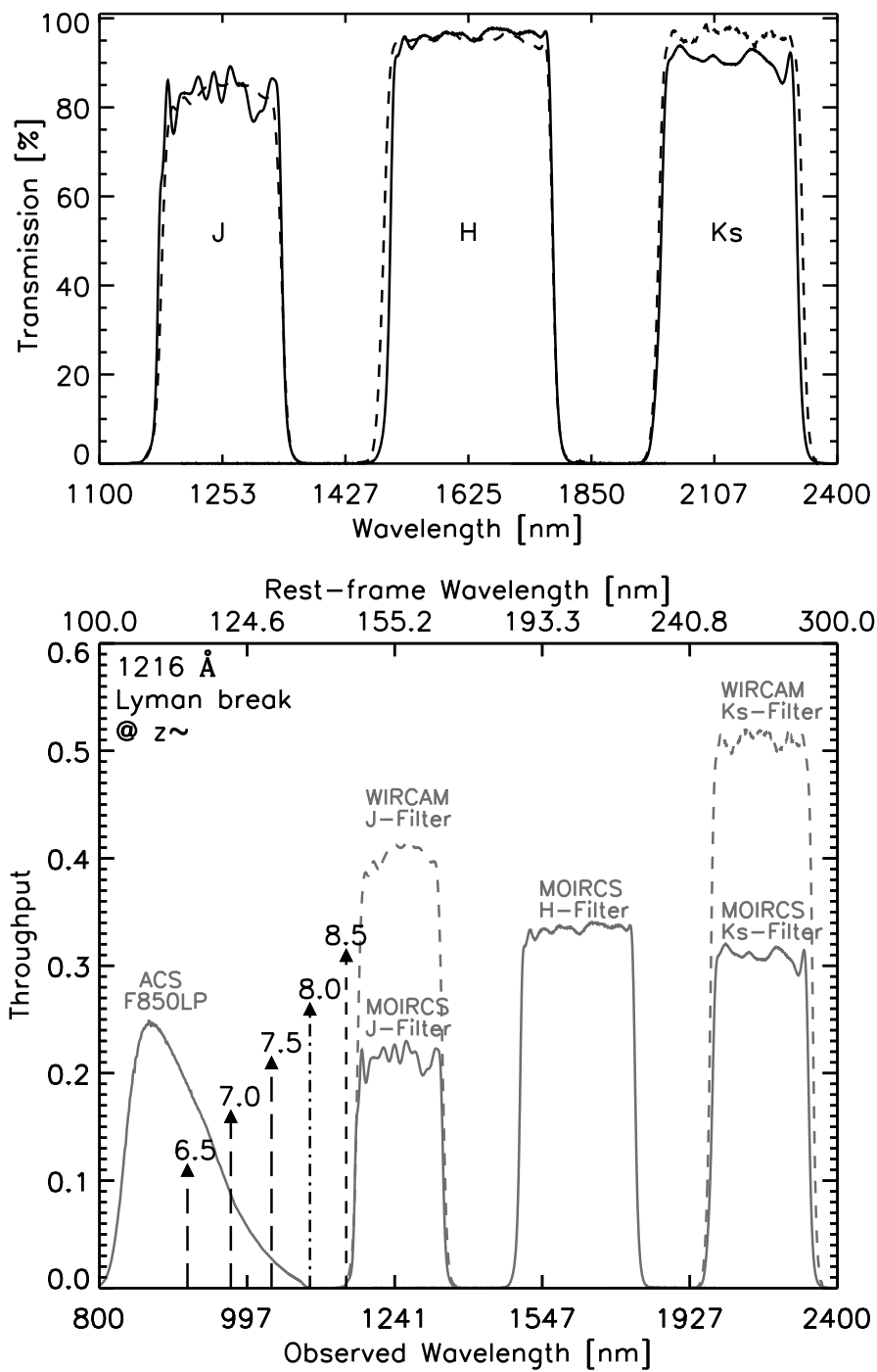}
\caption{[Top] The Subaru/\moircs\ \jm, \hm, \ks\ (solid) and the
  CFHT/\wcam\ \jw, \ks\ (dashed) filter transmission
  curves. [Bottom] Location of rest-frame 1216~\AA\ Lyman break at various
  redshifts ($z\gtrsim 6.5$) and filter throughput curves of
  \acsz-, $J$-, $H$-, \ks-bands. The top axis shows corresponding
  rest-frame wavelengths at $z\simeq 7$.}\label{fig:filters}
\end{center}
\end{figure}

The \wcam\ is on the CFHT 3.6m telescope and contains 4
2048$\times$2048 pixel HAWAII2-RG detectors, covering a total area of
20$\times$20 \sqarc\ with a sampling of
0.3\arcsec\ pixel$^{-1}$. \wcam\ is a near-infrared instrument
spanning the range 0.9--2.4 microns. Details of this instrument are
given in \citet{puge04}.  We used the CFHT archive
CADC\footnote{http://cadcwww.dao.nrc.ca/cadc/} to retrieve the
\wcam\ observations in the GOODS-N field. These observations were
carried out over a period of three years between 2006--2008
\citep[e.g.,][]{wang10}. They cover an area of about 1040~\sqarc,
and overlap completely with the GOODS-N ACS field as shown by a large
dot-dash box in \figref{fig:layout}. The typical seeing
for these observations is between 0.6\arcsec\ and 0.8\arcsec. These
observations were done in two broad-band filters \jm\ and \ks, whose
transmission curves are shown as dash curves in the top panel of the
\figref{fig:filters}. \tabref{tab:data} lists relevant information
about the NIR data in the GOODS-N field, used in this paper.

Filter transmission curves and total system throughputs used for these
observations are shown in \figref{fig:filters}. \wcam\ system
throughputs are calculated using their transmission
coefficients\footnote{http://www.cfht.hawaii.edu/Instruments/Imaging/WIRCam/WIRCamThroughput.html},
and \moircs\ system throughputs are estimated by scaling their
transmission curves to the peak throughput given in Table~1 of
\citet{ichi06}.

\subsection{\emph{HST}/ACS Observations}

The GOODS-North field was observed with \emph{HST}/ACS broad-band filters
\acsb, \acsv, \acsi, and \acsz\ \citep{giav04a}. The ACS imaging area
covers about 169~\sqarc\ as shown in \figref{fig:layout}. We used the
GOODS team reduction of the ACS images \citep{giav04a}. These images
have been drizzled from the original ACS pixel scale of
0.05\arcsec\ on to a grid of 0.03\arcsec\ pixels. We used v2.0 of the
ACS GOODS images\footnote{http://archive.stsci.edu/pub/hlsp/goods/v2/}
and corresponding AB magnitude zeropoints.  The combination of
deep \emph{HST}/ACS \bviz\ and ground-based $J$, \ks\ observations provide
the wavelength coverage needed for selecting LBGs at
\zdrop\ (\figref{fig:filters}).

\section{Near-IR Data Reduction}\label{data}

We have reduced publicly available \wcam\ and \moircs\ NIR imaging
data in the GOODS-N field by using a reduction pipeline based on the 
Interactive Data Language (IDL) --- Simple Imaging and Mosaicking PipeLinE
(\texttt{SIMPLE}) developed by Wei-Hao Wang.  Details of this
pipeline and the steps required to reduce the \wcam\ data
are given in \citet{wang10}.  We have independently reduced the \wcam\
data using this pipeline. Reducing the \moircs\ data is less straightforward and 
we took the following steps, using the \texttt{SIMPLE} software. Many general
steps are similar for both datasets.

\begin{itemize}

\item Generate File lists: We selected all archival observations taken
  on \emph{photometric nights only}. SIMPLE requires input lists of the raw
  FITS files which are grouped in such a way that each group has same
  exposure time, is from the same chip, is taken on the same day, and
  has same coordinates. If a large group of exposures is taken at the
  same pointing then we break them up into smaller chunks.

\item Reduction: For the first pass reduction, these file lists are
  given as input to the main IDL procedure \texttt{reduce\_moircs.pro}
  included in the \texttt{SIMPLE} software. This script does basic
  reduction steps including the flat-fielding, cosmic ray removal,
  sky-subtraction, aligning images, distortion corrections, and
  general plate solution. It also produces exposure time maps.

\item De-fringing: The output of the first pass reduction is carefully
  checked for image defects and/or fringes. The \moircs\ data show
  strong circular fringes in both chips but more prominently in
  chip-2.  We use the program \texttt{defringe\_moircs.pro} provided in the
  SIMPLE package to remove the fringes. The program first masks
  detected objects in a flattened and sky-subtracted image and
  transforms the image to a polar coordinate system, where the
  circular fringes become nearly straight lines. Then it fits a 5th
  degree polynomial to each column along the fringes to produce a
  model fringe image. The fringe model is then transformed back to the
  original Cartesian coordinate and subtracted from the image.
  This procedure considerably improves the quality of the reduced
  images.  A very similar fringe removal procedure is adopted by the
  MOIRCS Deep Survey team \citep{kaji11}.  A few exposures with very
  bad fringes, which could not be removed through above mentioned
  process, and/or image defects are not included in the second pass
  reduction. The reduced images and the exposure time maps are used to
  form large mosaics.

\item Mosaics: The final combined mosaic is generated by the procedure
  \texttt{mosaic\_wide.pro}. 
  The final \moircs\ images in three filters are registered to each
  other and to the GOODS ACS images by making the WCS of the \moircs\
  images consistent with other multi-wavelength images of the GOODS
  survey. The final pixel scale for the \moircs\ images is
  0.15\arcsec\ pixel$^{-1}$. For proper comparison with the \wcam\
  0.30\arcsec\ pixel$^{-1}$ images, we also generated 0.30\arcsec\
  pixel$^{-1}$ \moircs\ images using the \texttt{IRAF} task `blkavg'.

\item Zeropoints: The procedure \texttt{reduce\_std\_moircs.pro} was
  used along with the standard star images from the archive to reduce
  standard star images and calculate zeropoints in all the three
  \moircs\ filters ($J$, $H$, \ks).  The \moircs\ images are in
  ADU/sec and their derived AB zero-points are 26.175~mag for \ks,
  26.569~mag for \hm\ and 25.965~mag for \jm\ filters. The \wcam\
  images are in $\mu$Jy units with a zeropoint of 23.9~mag.

\item Combined Images: We have generated combined (\wcam\ and \moircs)
  $J$, \ks\ images, which have similar seeing/point spread functions
  (PSFs), using the Terapix \texttt{SWarp} package \citep{bert02},
  which resamples and co-adds FITS images using astrometric projection
  defined in the WCS. The goal of these combined images is to generate
  the deep $J$ and \ks-images in the central $\sim$60\% of the GOODS-N
  field. The combination was performed as follows: First the images
  were scaled to a common flux scale. Absolute root mean squared (RMS)
  noise maps were generated for each mosaic by scaling the mean
  variance of the images to the mean inverse exposure time maps
  generated by the \texttt{SIMPLE} reduction package.  The data were
  then combined using a weighted mean combination with the
  \texttt{SWarp} package using the RMS images as the weight image.  A
  bi-linear interpolation kernel was used to re-sample the images onto
  a common pixel grid. The final combined $J$, and \ks\ images are in
  $\mu$Jy units and have a zeropoint of 23.9~mag.

\end{itemize}

\section{Near-IR Data Quality and Catalogs}\label{quality}

We have reduced archival $J$, \ks-band \wcam\ and \moircs\ images, and
shallow $H$-band \moircs\ image of the GOODS-N field.  We then
generated combined images (\wcam\ and \moircs) in the $J$ and
\ks-filters, which covers much larger area than the \emph{HST}
data. \tabref{tab:data} shows area coverage information
of the NIR data in the GOODS-N field.  In this
section, we discuss the quality of these combined images, source
detections, and the generation of catalogs covering the entire area of
this field.

\subsection{Astrometry}

\begin{figure}
\begin{center}
\epsscale{1.0}
\plotone{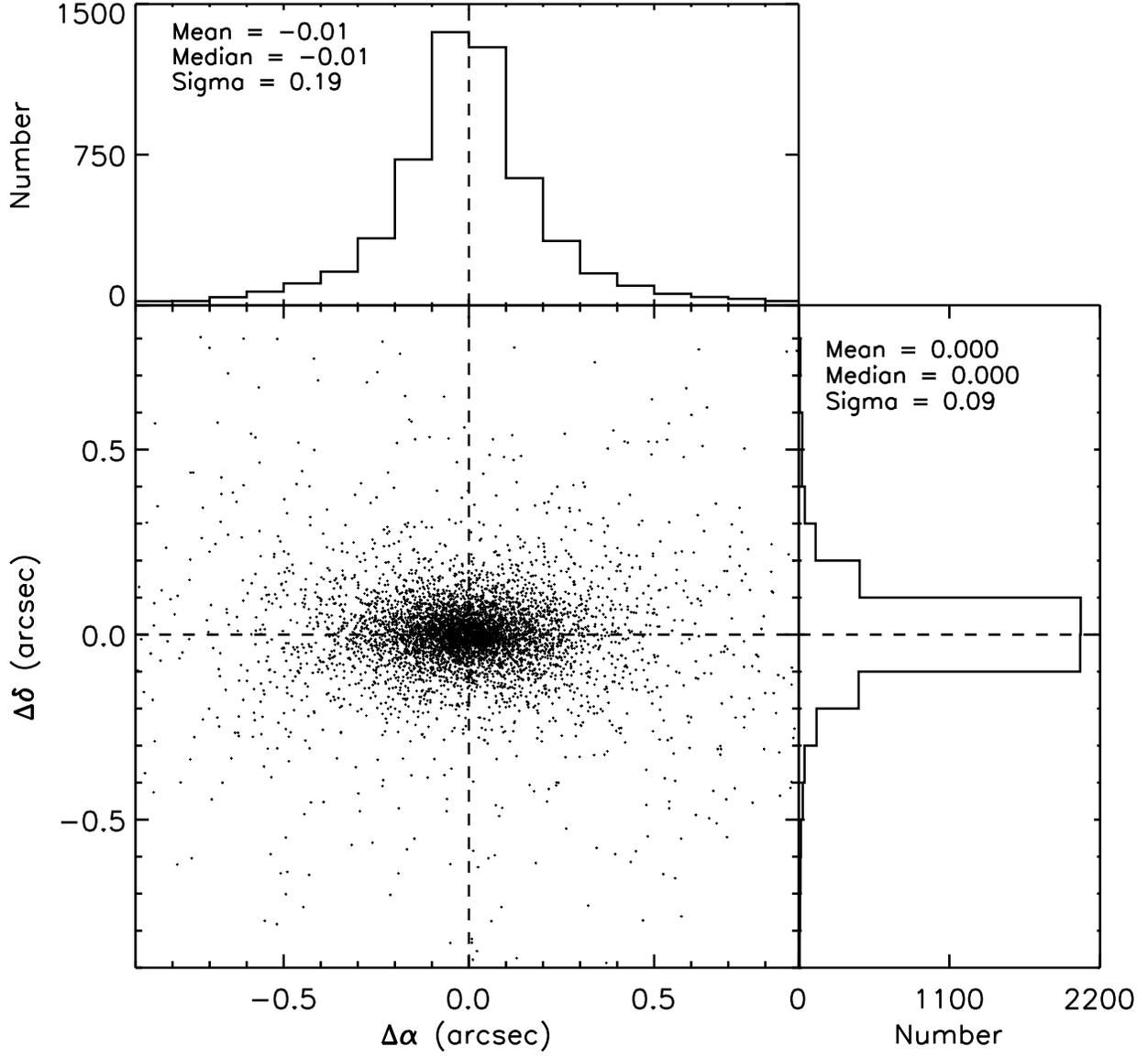} 
\caption{Relative astrometric offsets between the combined
  \ks-selected and the GOODS-N ACS \acsz-selected catalogs. Compact
  sources with \sn\ $>$ 20 and FWHM $<$ 1.2\arcsec\ are plotted. The
  histograms show distribution of the offsets in RA and
  DEC.}\label{fig:astro}
\end{center}
\end{figure}

The astrometric solutions of the final mosaiced images were tested by
comparing them with the available \emph{HST}/ACS \acsz-band catalogs.
\figref{fig:astro} shows the relative astrometric offsets in RA and
DEC between the \ks-selected and the GOODS-N ACS \acsz-selected
catalogs. For the purpose of this comparison, we selected well
detected/non-saturated compact sources with \sn\ $>$ 20 and full width
half maximum (FWHM) $<$ 1.2\arcsec\ from the GOODS-N ACS
\acsz-selected
catalog\footnote{http://archive.stsci.edu/pub/hlsp/goods/catalog\_r2/}
and matched them with the \ks-selected catalog from the combined
image. \figref{fig:astro} shows the distribution of the offsets with
their mean, median and sigma values. The uncertainty ($\sigma$) in the
offsets between the combined \ks-selected and the ACS \acsz-band
catalogs is 0.19\arcsec\ for RA (uncertainty in RA$\cdot$cos(DEC) is
0.09) and 0.09\arcsec\ for DEC. The uncertainty in the RA is slightly
higher than the DEC, but these are still less than one pixel in the
\ks-band (0.30\arcsec\ pixel$^{-1}$). These uncertainties include both
the internal errors in the ACS \acsz-band catalog and the errors
introduced during the image registration of SIMPLE. There are no
systematic offsets between the \ks\ positions and the ACS \acsz\
catalog.

\subsection{Photometry}

\begin{figure}
\begin{center}
\epsscale{1.0}
\plotone{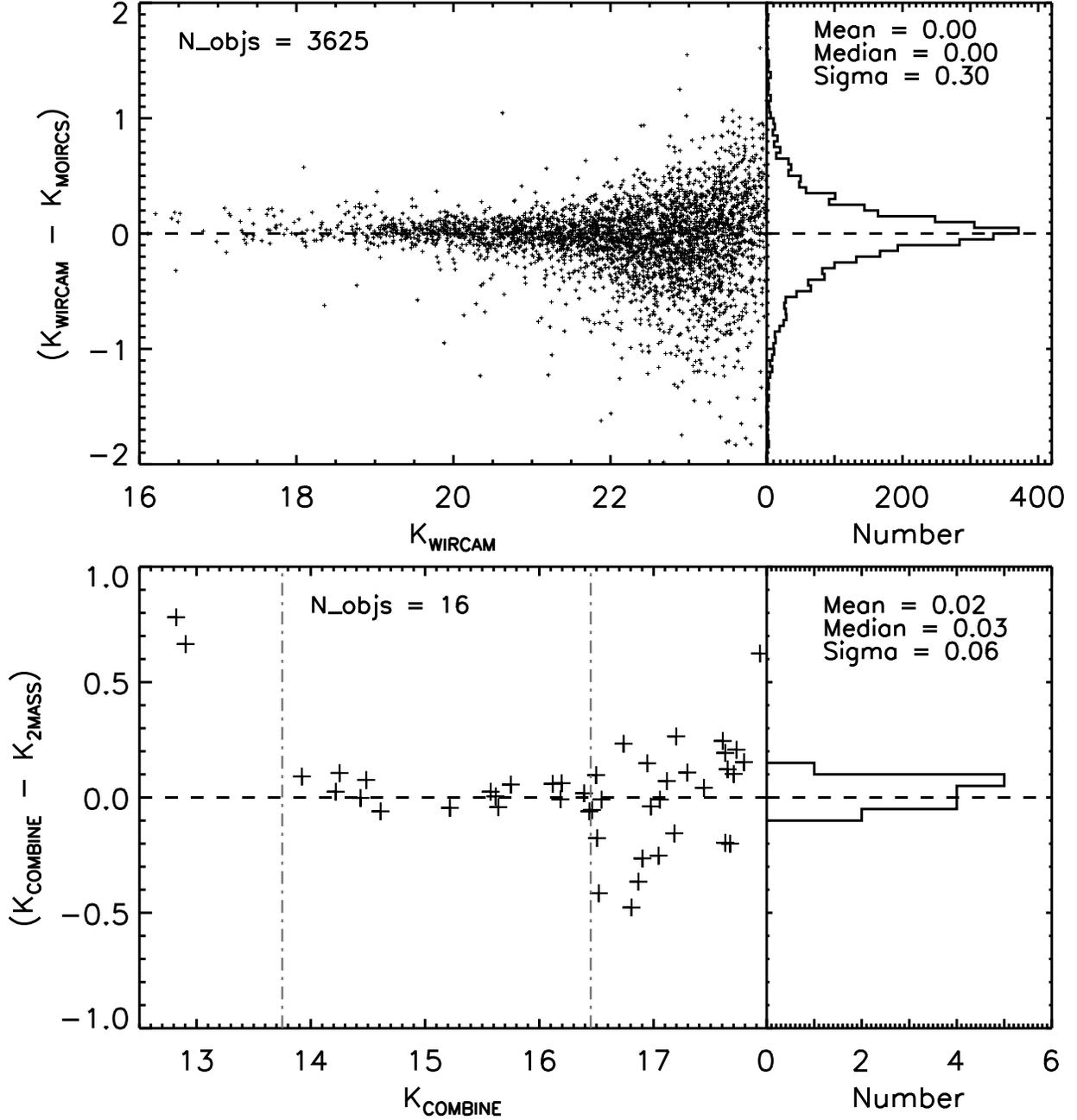} 
\caption{[Top] Comparison between  \ks-band magnitudes
  (\texttt{SExtractor} AUTO) obtained from the \wcam\ and the
  \moircs\ images. [Bottom] Comparison between \ks\ magnitudes
  obtained from the combined image and \ks\ magnitudes from the
  2MASS catalog. We compare 2MASS magnitudes in a magnitude range
  between $\sim$14 and $\sim$16~mag (dot-dash vertical lines in the
  bottom panel) because of the non-linearity/selection issues outside this
  magnitude range \citep[see][for details]{wang10}. }\label{fig:mags}
\end{center}
\end{figure}

To check the consistency of the photometry in our reduced images, we
compared the \ks-band magnitudes obtained using the
\texttt{SExtractor} \citep{bert96} AUTO apertures in the \wcam\ image
with the \moircs\ image, and also compared \ks-band magnitudes in the
combined image with the \ks-band magnitudes from the Two Micron All
Sky Survey (2MASS). The 2MASS magnitudes have been converted to the AB
magnitude system using the conversion given by \citet{cili05}.  The
top panel of \figref{fig:mags} shows the comparison between the
\texttt{SExtractor} \ks\ magnitudes, for all matched objects, from the
\wcam\ and the \moircs\ images.  There is no significant systematic
offset between these two magnitudes, and the mean of their difference
is $\sim$0~mag, with the uncertainty in the difference being
$\sim$0.30~mag.  The mean value (uncertainty) for
[\ks(\wcam)--\ks(\moircs)] is 0.03 (0.08) for
brighter objects (\ks$\,\lesssim\,$20~mag). The bottom panel of \figref{fig:mags}
shows the comparison between the \texttt{SExtractor} \ks\
magnitudes, measured in the combined images, and the \ks\ magnitudes
from the 2MASS catalog \citep{cutr03}. We compare these magnitudes
over the range $\sim$14 to $\sim$16~mag (as shown by dot-dash vertical
lines on the bottom panel of the \figref{fig:mags}) because of the
non-linearity issues in the \wcam\ and the selection effects
in 2MASS outside this magnitude range \citep[see][for details]{wang10}. The
comparison here is based on a small number of bright sources  and shows
no major systematic offset, with the uncertainty in the difference
being very small and mean magnitude difference of 0.02~mag.

\subsection{Catalogs and Number Counts}\label{counts}

\begin{figure}
\begin{center}
\epsscale{1.0}
\plotone{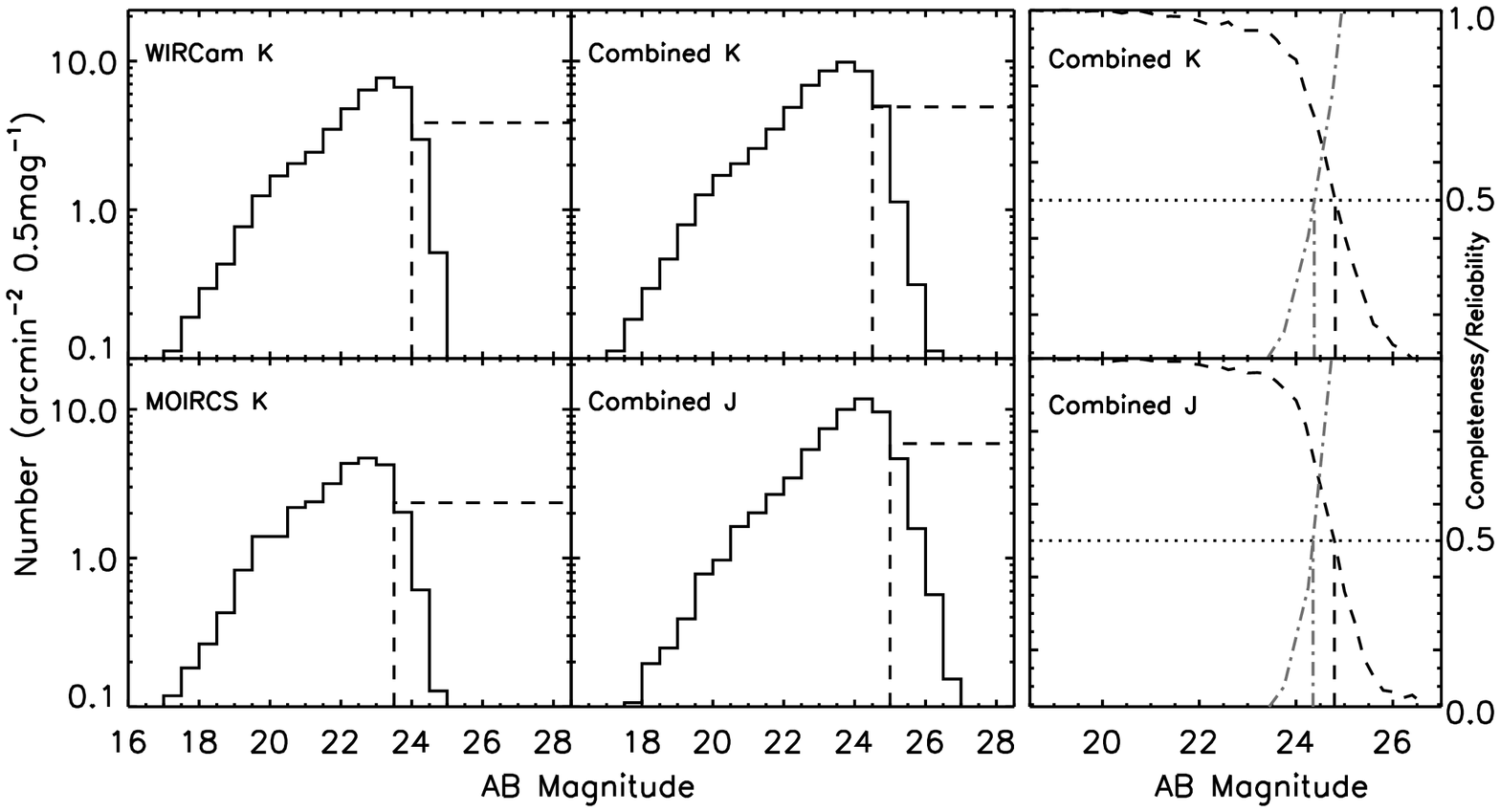}
\caption{ [Left 4-panels] The \ks-band number counts for each
  image. The bottom-left panels shows \moircs\ \ks-band number counts,
  while the top-left panel is for the \wcam. The bottom-right panel shows
  $J$-band number counts obtained from the combined image, while
  the top-right panel is for \ks-band. In all panels, the dashed
  vertical line shows the magnitude at which the number counts fall to
  50\% of their peak value. It is clear that combined images are
  $\sim$0.5~mag deeper than the \wcam\ or \moircs\ images. [Right
  2-panels] The \ks-band completeness obtained from the
  simulations. Details of these simulations are discussed in text
  (\secref{counts}). Top panel shows the completeness from the combined
  \ks-band image, while the bottom panel is for the combined $J$. The
  dashed vertical line shows the magnitude at which the recovery rate
  falls to 50\%. The dot-dash curve 
shows the reliability curve (\secref{counts}) for both $J$- and
\ks-selected catalogs. The dot-dash vertical line shows the magnitude
at which the reliability falls to 50\%.
 }\label{fig:ncounts}
\end{center}
\end{figure}

We used the \texttt{SExtractor} in dual-image mode with the \ks\ image
as the detection image for \ks-selected catalog, and $J$ image for
$J$-selected catalog. We generated the RMS maps from the
weight/exposure time maps and used them during the \texttt{SExtractor}
runs to get a better estimate of the photometric errors, and hence, of
the \sn. Here, \sn\ is defined as 1.0857 divided by the SExtractor
error in the total magnitude, which is similar to
FLUX\_AUTO/FLUXERR\_AUTO.  These RMS maps also help to exclude false
and spurious detections at the edges with low signal-to-noise
values. We set the DEBLEND\_MINCONT parameter to 0.0001,
DETECT\_MINAREA to 4 pixels (2 pixels for $J$-selected catalog) and
detection threshold to 2.0$\sigma$ (1$\sigma$ for $J$-selected) , and
used convolution with a 3 pixel Gaussian filter in the
\texttt{SExtractor} configuration file.  \tabref{tab:param} lists the
\texttt{SExtractor} parameters used to generate \ks\ and $J$-band
selected catalogs in the GOODS-N field.  We used \texttt{SExtractor}
MAG\_AUTO as a measure of the total magnitude.  The left 4-panels in
\figref{fig:ncounts} show the source counts (i.e., number per \sqarc\
per 0.5 mag bin) in \ks\ and $J$ bands. The leftmost two panels show
the \moircs\ and \wcam\ \ks-band counts. They both reach comparable
depths marked by dashed vertical lines. The right panels show the \ks-
and $J$-band number counts obtained from the combined image. Because
of the non-uniform exposure time in the \wcam\ and \moircs\ images,
the combined \ks-band image is $\sim$0.3--0.5~mag deeper than either
\wcam\ or \moircs\ image.

We quantify the depth and completeness in our \ks- and $J$-band images
by inserting numerous fake sources of varying magnitudes into the
reduced images and recovering them using the same \texttt{SExtractor}
parameters (\tabref{tab:param}) used for the real sources. Fake
sources were given Gaussian profiles with the FWHM values similar to
the point sources in our image. We define the image depth by the
magnitude corresponding to a recovery rate of 50\%. This depth implies
a slightly lower detection rate for real galaxies, which are more
extended. The rightmost two panels in \figref{fig:ncounts} show the
3$\sigma$ completeness curves (dashed curves) for both images. The top
panel shows completeness for \ks-band image, while the bottom panel is
for the $J$-band.  The dashed vertical line in each panel shows the
magnitude at which the number counts fall to 50\% of their peak value,
which we consider as a measure of the completeness limit for these
images.

We also quantify reliability of our catalogs by doing the negative
image test. We multiply the science (positive) image by --1 to get a
negative image. We then apply same detection procedure using
\texttt{SExtractor} parameters listed in \tabref{tab:param}.  The
number of detected `objects' in these negative images quantify
reliability of these catalogs. The dot-dash curve in the rightmost two
panels of \figref{fig:ncounts} shows the reliability curve for both
$J$- and \ks-selected catalogs. The dot-dash vertical line shows the
magnitude at which the reliability falls to 50\%.

The simulations and the observed number counts show that both ($J$ and
\ks) combined images are complete to AB magnitude limit of
$\sim$25~mag (3$\sigma$), and 50\% reliable to $\sim$24.5~mag. We also
have $H$-band data in the GOODS-N region from the Subaru telescope
observations.  The $H$-band \moircs\ image is shallower compared
to $J$ or \ks-bands by $\sim$1~mag, but can be very useful for
brighter objects in this field.

\section{Lyman Break Galaxies at \zdrop\ (\acsz-dropouts)}\label{lbgs}

As an application of these NIR data in the GOODS-N field, we
combine these with the existing \emph{HST}/ACS data to look for
\acsz-dropout galaxies, which are LBG candidates at \zdrop.
To identify \acsz-dropouts, we search for Lyman-break signature that
occurs at the rest-frame 1216~\AA\ \citep{mada95} using the dropout
technique. This technique is based on photometric color selection and
requires imaging in at least two filters, one to the blue side of the
break and the other to the red side (usually we have two or more
filters on both sides). The presence of Lyman break makes
high-redshift galaxies much fainter in the blue band than in the red
one, or in other words, it makes them appear to ``drop-out" from the
blue band. For this reason, this method is also known as the dropout
selection, and the candidates found in this way are generally referred
to as ``dropouts". At $z\gtrsim 6.5$, this break moves through
observed \acsz-band ($\sim$9100~\AA).  \figref{fig:filters} shows the
locations of the rest-frame 1216~\AA\ Lyman break at $z\gtrsim
6.5$. It is clear that ground-based $J$ and \ks\ band filters,
combined with the ACS bands are crucial in identifying LBG candidates
at \zdrop.

The drizzled ACS images in \acsb, \acsv, \acsi, \acsz\ filters have a
spatial resolution of 0.03\arcsec\ pixel$^{-1}$, compared to combined
$J$ and \ks-band images which have 0.30\arcsec\ pixel$^{-1}$
resolution.  Therefore, to perform matched aperture photometry, we
generated re-binned (10$\times$10), PSF matched, ACS images to
properly compare with combined $J$ and \ks-band images.  Details of
catalog generation, completeness, and reliability are discussed in
\secref{counts}. In summary, we performed matched-aperture photometry
on \bviz, $J$, \ks\ images by running the \texttt{SExtractor} in the
dual-image mode with $J$-band as the detection image. We used
corresponding RMS maps to reduce the number of spurious sources
detected in the low signal-to-noise ratio edge of the images.  For
object identification, we adopted a limit of at least 2 contiguous
pixels above a threshold of 1$\sigma$. We constructed a $J$-selected
catalog using \texttt{SExtractor} parameters shown in
\tabref{tab:param}. We used \texttt{SExtractor} MAG\_AUTO, with the
default Kron factor$=$2.5 and minimum radius$=$3.5, as a measure of
the total magnitude.  This catalog was used to select LBG candidates
at \zdrop\ based on the dropout color selection criteria. The colors
of the objects were estimated from the matched-aperture MAG\_AUTO
photometry on the \emph{HST}/ACS and combined NIR
images, as discussed above.

\subsection{\acsz-dropout Selection}\label{lbg_sample}

\subsubsection{Color Selection}\label{color}

\begin{figure}
\begin{center}
\epsscale{0.5}
\plotone{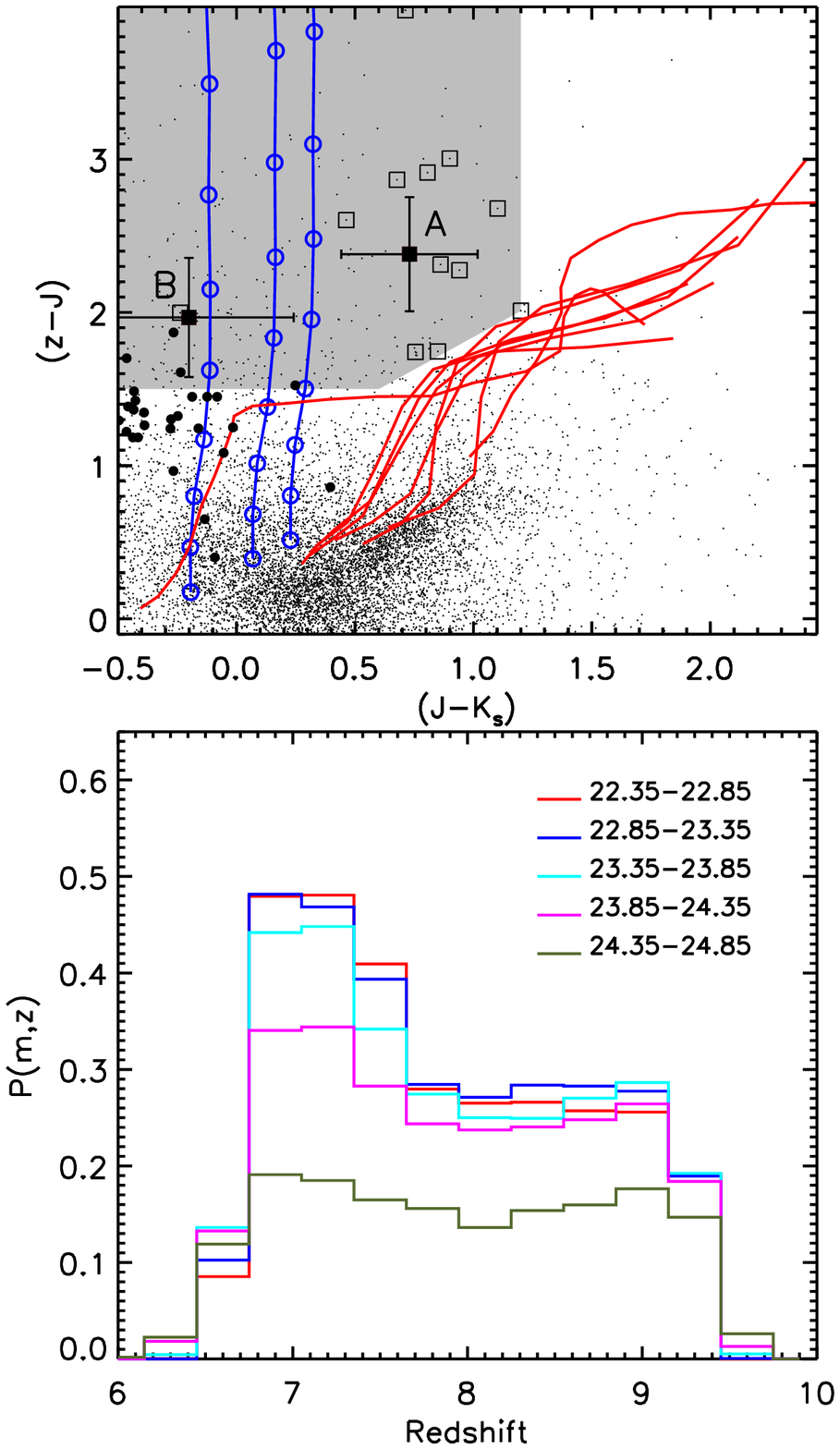} 
\caption{[Top] Selection of LBGs at \zdrop\ using ($z$--$J$) versus
  ($J$--\ks) colors. The gray shaded region is the selection
  region. The red curves are BC03, \citet{kinn96} and \citet{cole80}
  model colors of low redshift ellipticals, while blue curves (with
  open circles indicating different redshifts starting from the bottom
  $z=$6.0, 6.2, 6.4 ...)  are BC03 models of star-forming galaxies
  with E(B--V)=0, 0.15, 0.30~mag corresponding to three different
  curves. The black filled circles are expected colors of late-type
  stars based on \citet{pick98}.  The black data points (dots) are
  \emph{all} objects in the catalog. The filled black squares are two
  \acsz-dropout candidates, and the open black squares are 12 other
  candidates selected based on this color criteria before checking
  their \emph{Spitzer} magnitudes and photometric redshifts. The black
  data points (dots) in the selected region were excluded by other
  criteria as given in \secref{lbg_sample}. [Bottom] Redshift
  selection functions at different magnitude bins are shown by
  color-coded histograms.}\label{fig:lbgclr}

\end{center}
\end{figure}

The \acsz-dropout selection uses color criteria obtained from the
stellar population models of \citet[BC03]{bruz03}. \figref{fig:lbgclr}
(top panel) shows the BC03 star-forming galaxy models with E(B--V)=0,
0.15, 0.30~mag (solid blue lines with open circles showing different
redshifts), expected colors of late-type stars (black filled circles)
from \citet{pick98}, and tracks of lower-redshift ellipticals (red
lines) from three different models \citep[BC03,][]{kinn96,cole80}.
Our selection region is shown as a gray shaded area. We have corrected
the colors for IGM attenuation using the prescription of
\citet{mada95}. The two-color dropout selection criteria adopted here
are similar to those used to identify and study LBG candidates at
$z\simeq 1$--8 \citep[e.g.,][]{giav04b,bouw08,hath08,yan10,hath10}.
The $J$-selected catalog is used to select \acsz-dropouts from the
(\acsz\ -- $J$) vs. ($J$ -- \ks) color-color diagram
(\figref{fig:lbgclr}) using the following criteria:
\begin{displaymath}
      \left\{ \begin{array} {ll} (z_{\rm 850}-J) > 1.5 \:\hbox{mag}
        \\ \hbox{and}\: (J-K_{\rm s}) < 1.2 \:\hbox{mag}
        \\ \hbox{and}\: (z_{\rm 850}-J) > 0.99 + [0.85 \times (J-K_{\rm s})] \:\hbox{mag} 
        \\ \hbox{and}\: (J-K_{\rm s}) > -0.5 \:\hbox{mag}
        \\ \hbox{and}\: S/N (J) \ge 2.0 
         \end{array} \right.
\end{displaymath} 
We also require non-detection of the candidates (\sn$\,\lesssim\,$1)
in ACS optical \acsb, \acsv, \acsi-bands. The \sn\ cut in the
$J$-band ($\ge\,$2) corresponds to an AB magnitude limit of 
$\sim$25.3~mag. We have also applied the
additional criterion ($J$ -- \ks) $>$ --0.5 mag to eliminate the
possibility of selecting spurious candidates \citep[e.g., extreme
  equivalent width objects at lower redshifts:][]{vand11,atek11}, since this
color-color space is based only on \acsz, $J$, and \ks\ filters, and
it is required that LBG candidates be detected and are relatively bright
in the \ks-band.  After initial color and \sn\ cuts, we visually
inspected each selected object to remove any spurious source (due to
image defects or their proximity to a bright foreground object, edge
effects, or faint stellar diffraction spike).  This leaves 14
potential candidates as shown in the top panel
of the \figref{fig:lbgclr}.

\subsubsection{Redshift Selection Functions}\label{pmz}

To generate redshift selection functions, we ran simulations to
calculate $P(m,z)$, which is the probability that a galaxy with
apparent magnitude $m$ at redshift $z$ will be detected in the image
\emph{and} will meet our color selection criteria (\secref{color}). In
these simulations, large numbers of artificial objects with a range of
redshifts and magnitudes were added to real images, and then recovered
using exactly the same method and selection criteria that were
employed for the real observations. We simulated these objects in four
\emph{HST}/ACS, $J$, and \ks\ bands because we use these filters to perform
dropout color selection as shown in \secref{color}. The input spectrum for each
simulated object is from BC03 models with constant star-formation
history, three different E(B--V) values, two different metallicities,
and applying \citet{mada95} prescription for IGM attenuation below
rest 1216~\AA. The bottom panel of the \figref{fig:lbgclr} shows the
redshift selection function for different $J$ magnitudes.

\subsubsection{\emph{Spitzer}/IRAC Colors}\label{pmz}

The GOODS-N field has deep \emph{Spitzer}/IRAC imaging in [3.6],
[4.5], [5.8], and [8.0] channels. These observations are vital in
assessing the reliability of our \acsz-dropout candidates.  The ($J$
-- [3.6]) and ($J$ -- [4.5]) colors are used to differentiate between
extremely red objects \citep[EROs;][]{yan04} at $z\simeq 2$ and high
redshift candidates. The EROs have ($J$ -- [3.6]) and ($J$ -- [4.5])
colors in excess of $\sim$2.5 \citep{yan04}, putting a constraint on
the colors of high redshift galaxy candidates to have bluer colors
\citep[e.g.,][]{eyle07}.  To examine this, we used \emph{Spitzer}/IRAC
      [3.6] and [4.5] images to estimate magnitudes of 14 candidates
      in these bands.  We simply calculated IRAC total magnitudes from
      3\arcsec-diameter aperture magnitudes and used aperture
      corrections given by \citet{yan05}.  To confirm our photometry
      across different NIR passbands, we compared our
      \texttt{SExtractor} magnitudes with the Template FITting
      \citep[TFIT;][]{laid07} magnitudes.  The TFIT technique performs
      consistent multi-waveband photometry on images with widely
      different resolutions.  Our $J$, \ks, [3.6], and [4.5]
      \texttt{SExtractor} magnitudes are very similar (within
      uncertainties) to the TFIT magnitudes obtained using different
      set of proprietary CFHT NIR images in this region. To confirm
      our color selection and to differentiate between LBG candidates
      at \zdrop\ and EROs at $z\simeq 2$, we estimated photometric
      redshifts for 14 candidates.

\begin{figure}
\begin{center}
\epsscale{1.0}
\plotone{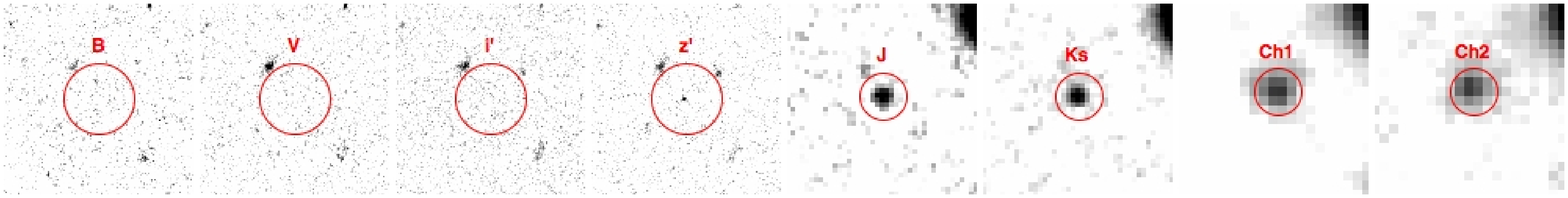}
\plotone{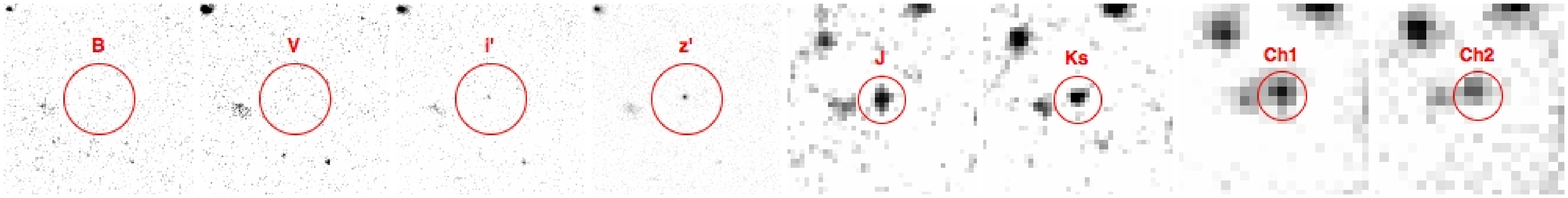}
\caption{Two possible candidates (top panel is candidate A, bottom
  panel is candidate B) at \zdrop. The coordinates and
  photometry of these candidates are shown in \tabref{tab:phot}. Left
  four stamps are ACS \bviz\ images at 0.06\arcsec\ pixel$^{-1}$,
  and right four stamps are lower resolution $J$, \ks, [3.6],
  [4.5] images, respectively. The circle shows the position of the
  object and is 1.5\arcsec\ in radius.}\label{fig:objs}
\end{center}
\end{figure}

\begin{figure}
\begin{center}
\epsscale{1.0}
\plottwo{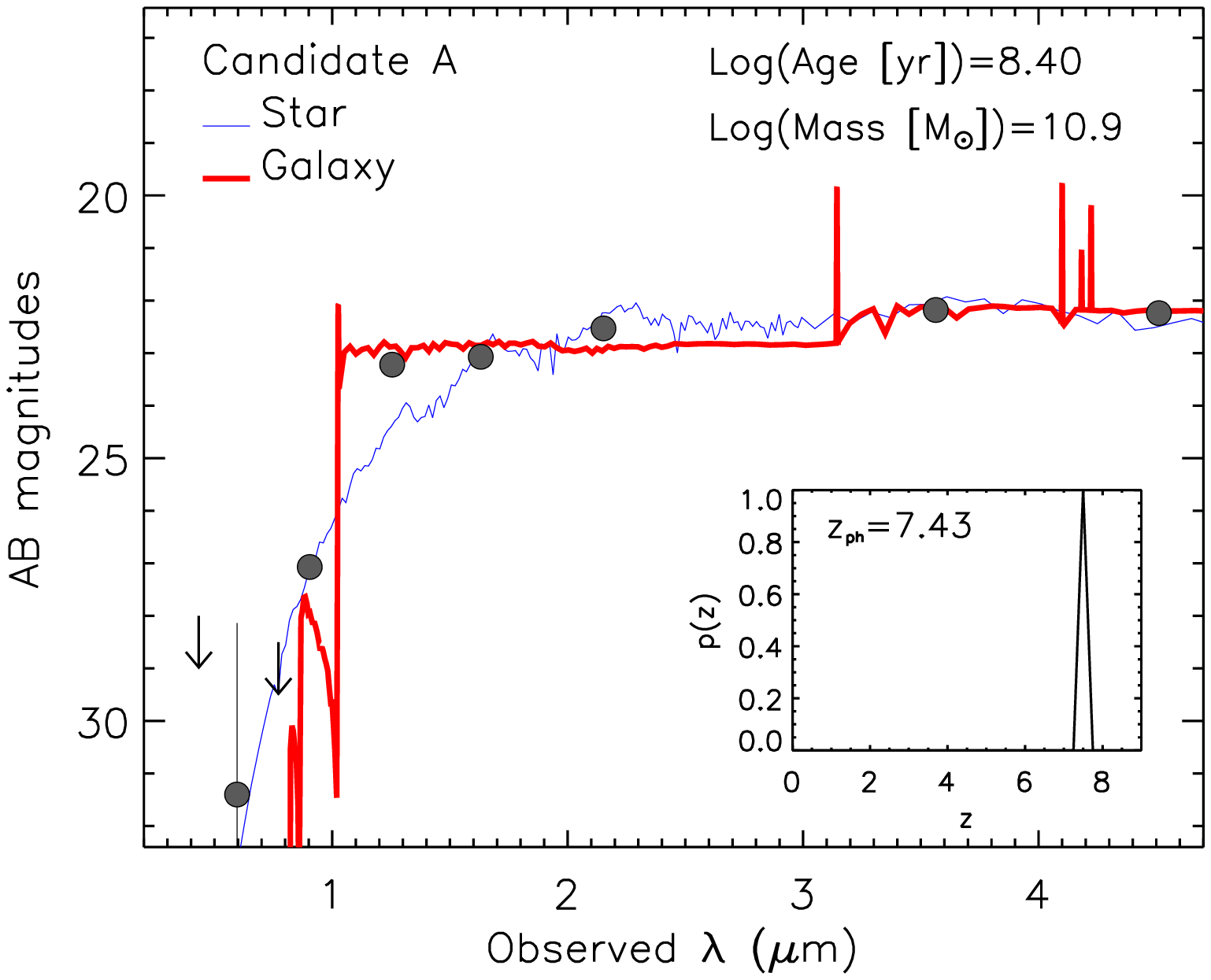}{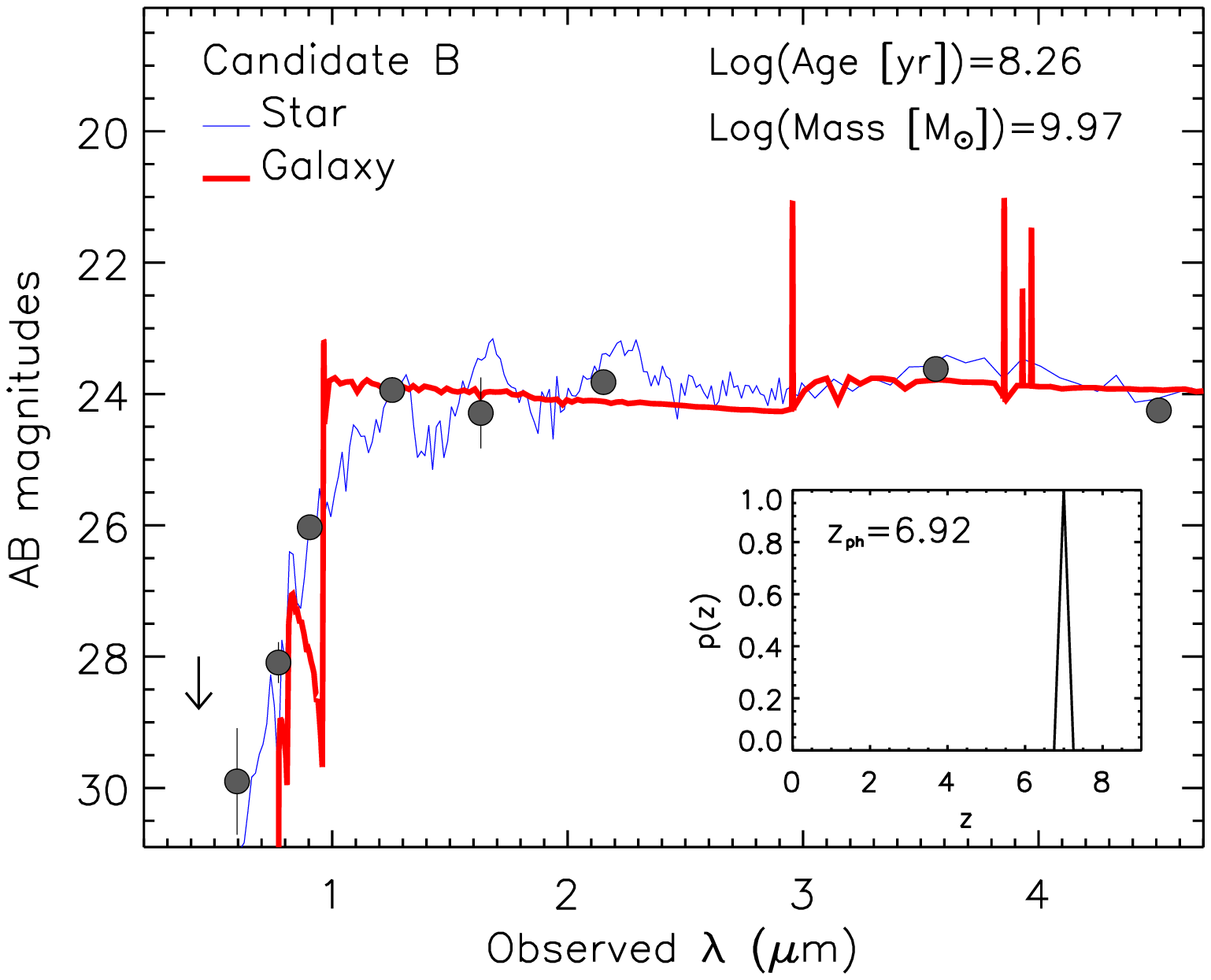}
\caption{Spectral energy distributions of two \acsz-dropouts.  The
  stellar SED (blue) and a galaxy SED (red) are shown for each
  object. Based on these SEDs, Candidate B is equally likely to be a
  star and a \acsz-dropout galaxy.  If these candidates are \acsz-dropout galaxies
  then we also show their best-fit stellar ages and stellar masses based on
  their SED fits.  We have used \texttt{Le PHARE}
  \citep{arno99,illb06} SED/photometric redshift code. The TFIT
  photometry (except \texttt{SExtractor} measured $H$~mag) is used for
  SED fitting because it is much more consistent for multi-wavelength
  data \citep[e.g.,][]{dahl10}.}\label{fig:seds}

\end{center}
\end{figure}

      We used three different codes to estimate photometric redshifts,
      namely \texttt{HyperZ} \citep{bolz00}, \texttt{EAZY}
      \citep{bram08}, and \texttt{Le
        PHARE}\footnote{http://www.cfht.hawaii.edu/$\sim$arnouts/LEPHARE/lephare.html}
      \citep{arno99,illb06}.  Out of 14 candidates, the photometric
      redshift estimate for 12 objects was not consistent between
      three codes with few objects showing possible detection in
      optical bands during visual inspection. Therefore, we find two
      possible LBG candidates at \zdrop\ after applying our color
      criteria, visually confirming non-detections in optical bands,
      and based on our photometric redshift estimates.  These two
      candidates are shown in \figref{fig:objs}. \tabref{tab:phot}
      lists the coordinates and TFIT photometry of these
      candidates. Their spectral energy distributions and best-fit
      models based on \texttt{Le PHARE} code are shown in
      \figref{fig:seds}. Though photometric redshifts show high
      probability that these are LBG candidates at \zdrop, the exact
      nature of these LBG candidates is unclear, and we will explore
      their properties and few possible scenarios.

\subsection{Are these luminous objects at \zdrop?}\label{z7real}

Based on the photometric data, it is unclear whether the objects found
here are genuinely at \zdrop\ or not. Both objects meet the color
selections suggested by previous studies \citep[e.g.,][]{bouw08,
  capa11}, but they are very bright ($J \lesssim 24$~mag) to be at
this redshift, specially in a comparatively smaller area
($\sim$169~\sqarc) of the GOODS-N field.  In this section, we discuss
properties of these candidates.

\subsubsection{Morphologies}\label{morph}

The GOODS-N catalog of \acsi-band
dropouts\footnote{http://archive.stsci.edu/pub/hlsp/dropouts/}
selected using criteria outlined in \citet{beck06} includes these two
candidates, but this catalog is only based on \emph{HST}/ACS \acsb, \acsv,
\acsi, \acsz\ magnitudes. Addition of $J$ and \ks\ observations help
to distinguish between an object at $z\simeq 6$ and \zdrop,
depending on the location of Lyman break in the \acsz-band.  Both
these candidates are very faint but compact in \acsz-band (stellarity
$>$ 0.5) so there is a strong possibility that these objects are faint
point sources (late-type stars). The $J$-band \texttt{SExtractor}
stellarity for candidate A is 0.54, while it is 0.13 for candidate B,
but the $J$-band PSF is less sharper than ACS so $J$-band stellarity
could be a less convincing measure of compactness for these
objects. Therefore, based on \acsz-band compactness measurements, it
is possible that these \acsz-dropout candidates could be stellar-like.

\subsubsection{Passive Galaxies}

One of the major interlopers in selecting high redshift dropout
galaxies is the lower redshift passively evolving galaxies. The
4000~\AA\ Balmer break in old galaxies at $z\simeq 2$ could
potentially be mistaken for a Lyman break at \zdrop. These passive
galaxies can be modelled with suitable set of spectral synthesis
models. We use the BC03 models to predict the colors of such
objects. \figref{fig:lbgclr} (top panel) shows predicted colors of
these old galaxies as a function of redshift through various
evolutionary tracks plotted in red. The color selection region of
\acsz-dropouts (gray shaded region in the top panel of
\figref{fig:lbgclr}) based on (\acsz\ -- $J$) and ($J$ -- \ks) avoids
these tracks, and hence, minimizes contamination of our sample by these
passive galaxies.  Also, the EROs at $z\simeq 2$ have very red ($J$ --
[3.6]) and ($J$ -- [4.5]) colors in excess of $\sim$2.5 \citep{yan04}
while our \acsz-dropout candidates have much bluer NIR colors, which
also suggests that these candidates are not likely EROs at $z\simeq
2$.

\subsubsection{Late-type Stars}

The NIR colors of Galactic cool stars are also very similar to the
high redshift dropout candidates and can mimic the Lyman break in the
color-color space as they are also optically faint and bright in the
NIR. The main distinguishing properties between a high redshift galaxy
candidate and a cool dwarf are their morphologies and NIR colors. We
discussed morphologies in \secref{morph}. Here, we investigate the
$J$, \ks\ and \emph{Spitzer}/IRAC colors which will shed light on the
nature of these \acsz-dropout candidates. The (\acsz\ -- $J$) and ($J$
-- \ks) colors in \figref{fig:lbgclr} show expected colors of
late-type stars (black filled circles) based on \citet{pick98}
models. The colors of candidate B fall in the region occupied by the
cool dwarfs, while candidate A has redder ($J$ -- \ks)
color. Therefore, based on these colors, it is possible that candidate
B could be a late-type star.

\citet{stan08} have proposed $J$, \ks\ and \emph{Spitzer}/IRAC color
cuts, which were designed to remove passive galaxies at $z\lesssim 2$
and late-type stars.  The \acsz-dropout candidate A successfully
passes these four color criteria, while candidate B does not pass the
([3.6] -- [4.5]) color cut. These criteria also indicate that the
candidate B could be an interloper (most likely a late-type star)
based on these NIR colors.

\subsubsection{Photometric Redshifts}

\citet{kaji11} have analysed \moircs\ data, separating it into two
regions --- the deep and the wide.  Our \acsz-dropout candidates are
not located in the deep \moircs\ region, but are covered by the
shallower wide region.  \citet{kaji11} estimated photometric redshifts
of these candidates using their own \moircs\ wide photometric
catalog. They measured photometric redshift using four different
techniques. Three out of four techniques estimate redshift of these
candidates to be at \zdrop, while the fourth approach (using
\texttt{EAZY} code) estimates lower redshift ($z\simeq 2$) for these
candidates.

We performed SED fitting using available multi-wavelength data
(optical to NIR) in the GOODS-N field to predict photometric redshifts
of our \acsz-dropout candidates.  We used three photometric redshift
codes (\texttt{Le PHARE}, \texttt{HyperZ} and \texttt{EAZY}) on TFIT
photometry to estimate their redshifts. The TFIT photometry generates
accurate photometric redshifts, as shown by \citet{dahl10} for the
GOODS-S field. These codes predict high probability for these two
candidates to be at \zdrop.  \figref{fig:seds} shows SED fitting using
\texttt{Le PHARE} code for two types of templates --- blue for a
late-type star (from \citealt{chab00} library) and red for a star
forming galaxy (from BC03 library) at \zdrop. The SED fitting for
candidate B shows equal probability (very similar $\chi^2$) for both
these templates, while candidate A has a higher probability for a LBG
at \zdrop. This implies that candidate B could be a late-type star,
but it is also equally likely to be a high redshift galaxy. We also note
that due to lack of $Y$-band data, the exact location of the Lyman
break between the \acsz- and $J$-band could be somewhat uncertain, and
that would increase the uncertainty in the photometric redshift.

\subsubsection{Keck Spectroscopy}\label{keck}

We used the DEep Imaging Multi-Object Spectrograph
\citep[DEIMOS;][]{fabe03} on the 10m Keck telescope to perform
spectroscopic observations of these two candidates during our
March--April 2011 observing run. The total exposure time for these
observations was $\sim$4 hours. The seeing was in the range of
0.5\arcsec\ -- 0.7\arcsec. We used the GC455 filter and the 600 line
mm$^{-1}$ grating, blazed at 7400~\AA.  The spectral coverage was
between $\sim$5500~\AA\ and $\sim$10000~\AA. The spatial pixel scale
was 0.1185\arcsec\ pix$^{-1}$, and the spectral dispersion was
0.65~\AA\ pix$^{-1}$. The slit widths were 1\arcsec.  The preliminary
reduction were performed using the \texttt{spec2d} IDL
pipeline\footnote{http://astro.berkeley.edu/\~cooper/deep/spec2d/}
developed by the DEEP2 Team \citep{coop12,newm12}.  Wavelength
calibration was performed by fitting to the arc lamp emission
lines. The standard stars and science targets were observed using the
same slit width so we did not apply any corrections due to slit-loss
effects.

\begin{figure}
\begin{center}
\epsscale{0.9} 
\plotone{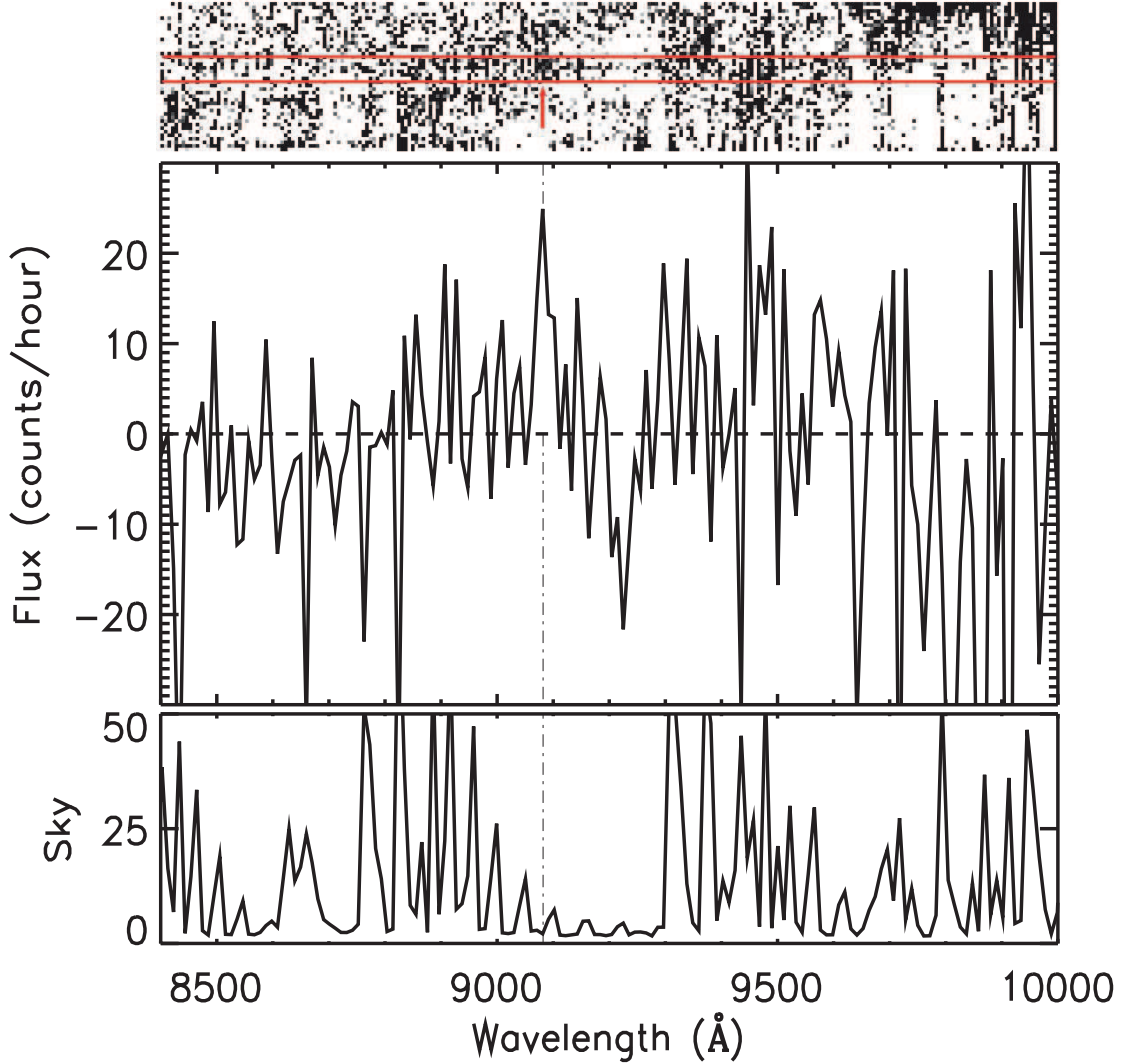}
\caption{[Top] Keck DEIMOS two dimensional spectrum of candidate B
    covering 8400--10000~\AA, displayed using histogram equalization, 
    is shown with the object position
    bracketed by horizontal red lines. The gray scale is such that the
    darker is more positive. [Bottom] Corresponding one
    dimensional spectrum obtain through 
    \texttt{spec2d} reduction software.
    There is a possible weak Ly$\alpha$ line detection at
    $\sim$9090~\AA\ in this spectrum, indicated by a dot-dash line in
    one dimensional and marked by a red arrow in the two dimensional
    spectrum. If confirmed, this \acsz-dropout candidate could be a LBG at
    $z\simeq 6.5$. }\label{fig:keck}
\end{center}
\end{figure}

We present the Keck spectrum of one \acsz-dropout candidate which has
high enough \sn\ ratio to make a redshift measurement possible. For
the second candidate, because of the unfortunate location on the mask
and sky lines, the spectrum is not reliable.  In \figref{fig:keck}, we
show one and two dimensional DEIMOS spectra of the \acsz-dropout
candidate B between 8400 and 10000~\AA.  \figref{fig:keck} shows a
weak emission line at $\sim$9090~\AA\ indicated by a dot-dash line in
the one dimensional spectrum, and by a red arrow in the two
dimensional stamp. The sky spectrum is shown at the bottom of the one
dimensional spectrum and there is no major sky line at this
wavelength. This spectral feature also shows asymmetric nature of a
Ly$\alpha$ line, and we don't see any clear detection of other lines,
so we tentatively identify this line as a Ly$\alpha$ at $z\simeq
6.5$. The estimated line flux is $\sim$1.4$\times$10$^{-17}$ erg
s$^{-1}$ cm$^{-2}$, and the  rest-frame equivalent width (EW) is
$\gtrsim\,$21~\AA. Hence, candidate B is a possible LBG at $z\simeq 6.5$,
consistent with its photometric redshift.  The red
part of the spectrum of candidate A is mostly affected by improper
masking and sky lines, so it is unreliable and inconclusive. The bluer
part of the spectrum (below 8000~\AA) does not show any
clear detection of emission line. This could imply that
candidate A is not a low redshift galaxy, but does not rule out other
possibilities (e.g., late-type star or LBG). This spectrum cannot
conclusively predict the nature of candidate A --- whose photometric
redshift is $\sim$7.4 --- because the possible Ly$\alpha$ emission
line would be around 10200~\AA, which we cannot identify. Therefore,
the Keck spectrum of candidate A cannot differentiate between a
late-type star or a LBG at \zdrop. We will try to re-observe these
candidates with Keck/DEIMOS or MOSFIRE NIR spectroscopy.

The three spectroscopically confirmed objects of \citet{ono12} at
\zdrop\ were also observed during the same Keck observing run. These
objects were about 1~mag fainter in the continuum than our candidates,
and all three showed detectable Ly$\alpha$ in emission. These
successful Ly$\alpha$ detections along with the likely Ly$\alpha$
detection in Candidate B allow us to put lower limits on
Ly$\alpha$ line emission from our \acsz-dropout candidate A. The three
\citet{ono12} objects have Ly$\alpha$ flux limits (1$\sigma$) in range
of 2.5--2.7$\times$10$^{-17}$ erg s$^{-1}$ cm$^{-2}$, and rest-frame
EW between 33 and 43~\AA. These flux and EW
measurements are consistent with other confirmed Ly$\alpha$ emitters
at \zdrop\ \citep[e.g.,][]{vanz11,pent11}. This implies that if our
candidate A is a Ly$\alpha$ emitter, then we could detect Ly$\alpha$
line with flux $\gtrsim\,$1.4$\times$10$^{-17}$ erg s$^{-1}$
cm$^{-2}$, and EW$\,\gtrsim\,$20~\AA.

The fraction of LBGs showing Ly$\alpha$ in emission is a strong
function of the absolute UV-magnitude (M$_{\rm UV}$), i.e., fraction
increases for fainter galaxies.  This has been confirmed by
observations at $z\simeq 6$--8 \citep[e.g.,][]{star11,sche12}, and
also by smoothed particle hydrodynamics simulations
\citep[e.g.,][]{fore12}. These studies imply that more luminous
LBGs are less likely to show Ly$\alpha$ in emission, and it is
possible that candidate A will not show Ly$\alpha$ line because of
its luminosity.  We should also emphasize that, at similar continuum
magnitudes, \citet{capa11} could not clearly detect Ly$\alpha$
emission for their LBG candidates at \zdrop.  Therefore, to clearly
identify the nature of our LBG candidates at \zdrop, we need deep
NIR spectroscopy from Keck/MOSFIRE or \emph{HST}/WFC3.

In this section, we investigated morphology, NIR/IRAC colors, and Keck
spectroscopy of our two \acsz-dropout candidates. We find that though
photometric redshift and NIR colors of candidate A support that it
could be a LBG at $z\simeq 7$, because of inconclusive Keck spectrum
the true nature of this candidate is uncertain. The candidate B could
be a real \acsz-dropout galaxy at $z\simeq 6.5$ based on a weak
Ly$\alpha$ detection in the Keck spectrum, but we cannot rule out
other possibilities based on our data.  Therefore, it is plausible
that one or none of these candidates are at \zdrop.  Hence, we
conclude that we find 1$\pm$1 LBG candidates at \zdrop, brighter than
AB$\,\sim\,$24.5~mag, in the GOODS-N field.

\subsection{Discussion}

\subsubsection{Other Surveys and Cosmic Variance}

The rarity of bright LBG candidates at \zdrop\ in a comparatively
smaller area ($\sim$169~\sqarc) questions as why other surveys have
not found such objects.  \citet{cons11} used 30 NICMOS pointings of
massive galaxies at $z\simeq 1.7$--$2.9$ to cover smaller area of the
GOODS-N field with NIR data.  These NICMOS observations were used by
\citet{bouw10b} to search for LBGs at \zdrop, but they do not cover
the area in which we find our two candidates.  The PEARS grism survey
\citep{malh07} in this field also does not cover this region, which
could have helped to confirm the nature of our \acsz-dropout
candidates.

\citet{ouch09} survey was a $Y$-band survey in the GOODS-N field and
does not use the NIR bands.  This survey requires a detection brighter
than 26~mag in the $Y$-band ($\sim$1 micron). Their redshift selection
function shows a survey optimized to select galaxy candidates at
$6.5<z<7.1$, therefore if our candidates are outside this redshift
range then \citet{ouch09} could not have found them.  Moreover, they
used Suprime-Cam on the Subaru Telescope to do a wide-area survey at 1
micron, which is not possible in the NIR simply because of the small
field of view of NIR detectors. As a result, they also surveyed the
extended GOODS-N region with no \emph{HST} data. In the present study,
we have generated deep NIR images in the GOODS-N area for which we
have very deep \emph{HST}/ACS and \emph{Spitzer}/IRAC data, all with
consistent photometry using TFIT technique. Combining these data with
the NIR data, we can then search for the \acsz-band dropout galaxies,
which puts limits on the source density of these sources. Also, the
availability of combined \ks-band data allows searches for high-$z$
galaxies when \emph{HST}/WFC3 ($YJH$) data become available through
the CANDELS \citep{grog11,koek11}.

Similar wide area NIR surveys have been done in the GOODS-S and other
fields to look for brighter LBG candidates at \zdrop.  \citet{hick10}
used ESO/VLT observations in the GOODS-S field covering
$\sim$119~\sqarc\ to search for high redshift LBGs. They found four
possible candidates at \zdrop\ with an AB magnitude brighter than
25.5~mag. \citet{hick10} survey covers slightly smaller area than our
survey and does not find any \acsz-dropouts with $J$ magnitude
brighter than 24~mag, but they do find objects with similar
\ks\ magnitudes as our candidates. \citet{cast10} also conducted a VLT
survey covering three different fields (total area 161~\sqarc) to
search for LBGs at \zdrop. They found a total of 15 candidates but all
are fainter than 25.5~mag. The \emph{HST}/WFC3 parallel imaging
surveys \citep{yan11,tren11} have covered $\sim$125~\sqarc\ area ---
which is smaller than our current survey area --- to search for
$Y_{098}$-dropouts, which is optimized to identify LBGs at $z\gtrsim
7.5$. These surveys have identified few candidates with AB magnitude
fainter than or equal to 25~mag. Though most of the surveys have
smaller area coverage than our full area GOODS-N field
($\sim$169~\sqarc), they don't find any LBG candidates at \zdrop\ with
magnitudes brighter than $J_{AB}\sim$24.5~mag.

\citet{stan08} performed extensive search for LBGs at \zdrop\ in
10 widely separated fields as a part of the ESO Remote Galaxy
Survey (ERGS). This survey covered $\sim$225~\sqarc\ and found one
possible LBG candidate at $z\gtrsim 6.5$ with $J_{AB}\sim$23.6~mag,
which is very similar to $J$ magnitude of our \acsz-dropout
candidates. If \citet{stan08} bright candidate is confirmed, then the
surface density of such bright LBG candidates at $z\gtrsim 6.5$ is
$\sim$0.4$\times$10$^{-2}$ per \sqarc\ in the ERGS.

Based on other wide area surveys, it seems very unlikely that both of
our bright candidates are truly at \zdrop. The GOODS-N field is a
comparatively small area to find two such bright LBG candidates at
such a high redshift. \citet{capa11} found three possible
\acsz-dropout candidates with similar brightnesses, but in a area as
large as 2~deg$^2$. That being said, if one of our candidates is
confirmed to be at \zdrop\ then the estimated surface density is
$\sim$0.5$\times$10$^{-2}$ per \sqarc, which is very similar to what
\citet{stan08} found in the ERGS. Therefore, finding such a bright LBG
candidate at \zdrop\ in a survey area of $\sim$169~\sqarc\ is a
statistical possibility based on cosmic variance.

\subsubsection{Implications}

We find two possible LBG candidates at \zdrop\ from our analysis of
NIR images in the GOODS-N field ($\sim$169~\sqarc). The first
candidate could be a real LBG at $z\simeq 6.5$ based on a possible
Ly$\alpha$ line in the Keck spectrum, and the photometric redshift of
the second candidate support its high redshift nature. If confirmed,
these are among the brightest LBG candidates at \zdrop. At a redshift
of $z\simeq 7$, we estimate absolute magnitudes M$_{\rm UV}
\simeq$--23.7 and --23.1~mag for candidates A and B, respectively,
which is $\sim$1~mag brighter than spectroscopically confirmed
Ly$\alpha$ emitting \acsz-dropouts by \citet{ono12}. The two possible
LBG candidates at \zdrop\ in the GOODS-N field implies a surface
density of $\sim$1.0$\times$10$^{-2}$ per \sqarc\ at $J <
24.5$~mag. This is consistent with other surveys
\citep[e.g.,][]{stan08,ouch09,yan11} focusing on galaxies brighter
than 25.5~mag.

Using \citet{mada98} formalism, the rest-frame UV luminosities of our
candidates correspond to star formation rates (SFRs) of
$\sim$100--200~M$_{\odot}$ yr$^{-1}$. As a rough estimate, if they
keep forming stars at the same rates to z$\,\sim\,$5, they could
accumulate stellar masses of (0.5--1.0)$\times$10$^{11}$~M$_{\odot}$
in next $\sim\,$400 million years. Therefore, our luminous
\acsz-dropout candidates could be progenitors of massive LBGs
($\sim\,$10$^{11}$~M$_{\odot}$) observed at z$\,\sim\,$5
\citep[e.g.,][]{wikl08}.

The bright-end of the luminosity function declines exponentially so we
expect fewer bright galaxies at these redshifts. The surface density,
which we predict based on our luminous LBG candidates at \zdrop, is
significantly higher than what is expected from smaller area fainter
\emph{HST} surveys, but such a difference at the bright end of the
luminosity function is not surprising and is within the theoretical
limits based on halo mass function \citep{bouw08,capa11}. The star
formation processes are not well understood at these redshifts, so it
is very likely that the mechanism involving truncation of star
formation and feedback are different at higher redshifts and this can
lead to a flatter bright-end for the luminosity function.

\begin{figure}
\begin{center}
\epsscale{1.0}
\plotone{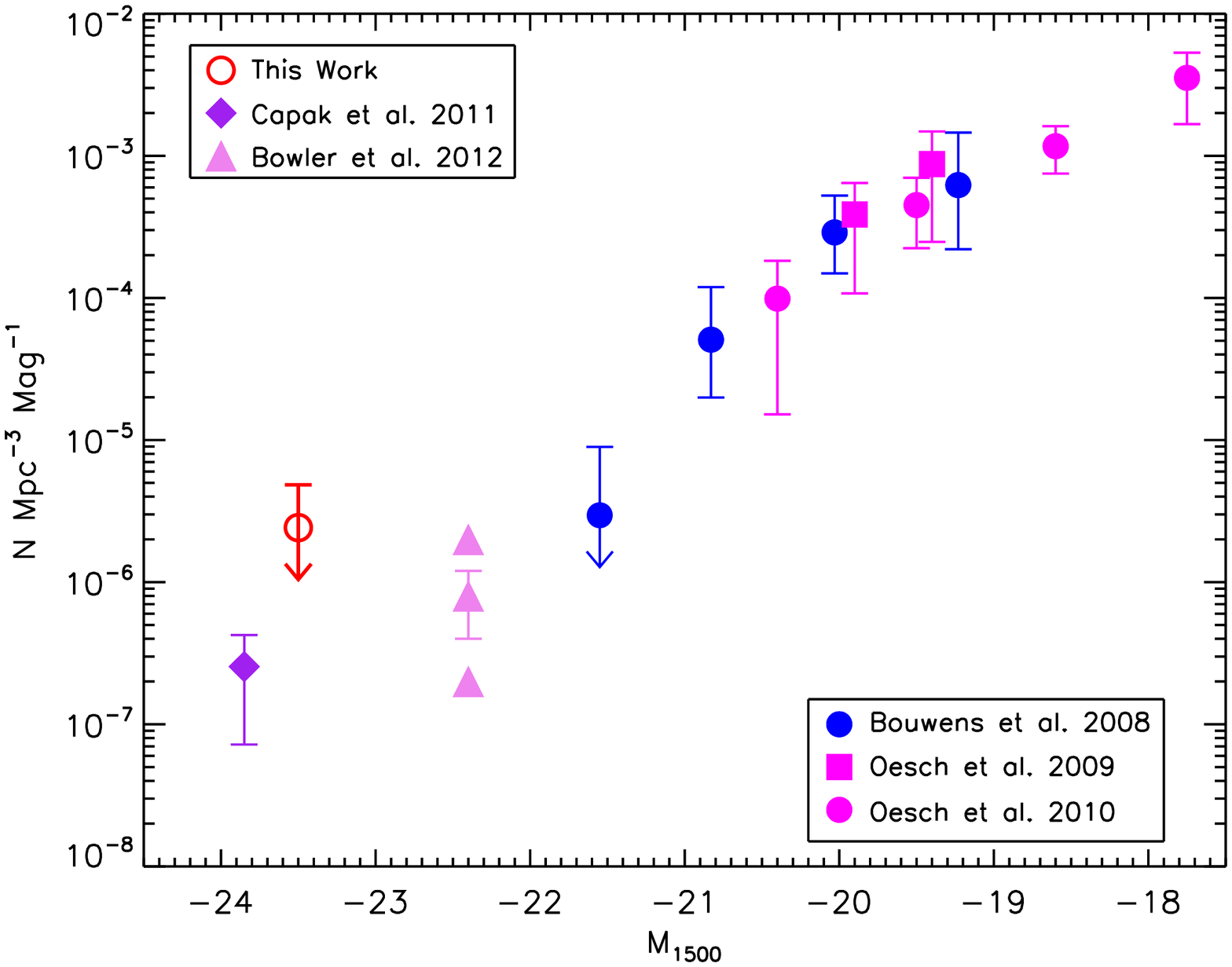}
\caption{Constraints on the bright-end of the luminosity function at
  \zdrop\ \citep{bouw08,oesc09,oesc10,capa11,bowl12} along with our number
  density based on one possible candidate  plotted as open
  red circle. The upper limit is for two possible candidates, while lower
  limit (down arrow) means no \zdrop\ candidates in
  this area at this brightness, and the number density has to be less than this
  value. The small area-deep surveys are not
  sufficient to probe the bright-end of the luminosity function.  We
  will need wide area surveys \citep[e.g.,][]{tren11,yan11,grog11} to
  explore these luminous galaxies at these
  redshifts.}\label{fig:numdens}
\end{center}
\end{figure}

To estimate the number density of these possible bright
\acsz-dropouts, the GOODS-N volume was estimated using $P(m,z)$
simulations (details in \secref{pmz}). This volume estimate is very
similar to the estimate obtained using cosmic variance calculator of
\citet{tren08}. The number density for one object at \zdrop\ obtained
using this volume is plotted in \figref{fig:numdens} along with other
number densities from the literature
\citep{bouw08,oesc09,oesc10,capa11,bowl12}.  The number densities at M$_{\rm
  UV} \simeq$--23.5~mag, if confirmed, point to a `no-knee' type
luminosity function at \zdrop, which has been considered as a
possibility for $z>6$ luminosity functions
\citep[e.g.,][]{bouw08,capa11}. The bright-end of the luminosity
function cannot be constrained by faint, small area surveys. To find
brightest galaxies and get better constraints for the bright-end of
the LF at these redshifts we will need wide area, shallow surveys
\citep[e.g.,][]{tren11,yan11,grog11}.  Though these bright objects are
luminous, it is still difficult to get spectroscopic confirmations ---
even with 10m Keck telescope (as discussed in \secref{keck}) --- if
they are not strong Ly-$\alpha$ emitters \citep[also
see][]{capa11}. The high resolution \emph{HST} imaging in NIR with the
CANDELS and deep WFC3/IR grism observations \citep[e.g.,][]{bram12} may help to confirm these
and other such candidates.

\section{Summary}\label{summary}

We have combined archival NIR images obtained from the Subaru and the
CFHT telescopes to generate deep $J$ and \ks\ band images in the
GOODS-N field. These NIR data are vital for variety of
multi-wavelength science goals. We use these images to search for
bright \acsz-dropouts, (i.e., LBG candidates at \zdrop). We find two
likely candidates at \zdrop. We attempted Keck/DEIMOS spectroscopy to
identify true nature of these candidates. One candidate has a weak
spectral feature which we tentatively identify as Ly$\alpha$ at
$z\simeq 6.5$. Though the colors and photometric redshifts of the
second candidate indicate that it is likely to be at \zdrop, the true
nature of this candidate is ambiguous. Hence, based on our analysis,
we predict 1$\pm$1 LBG candidates at \zdrop, brighter than
AB$\,\sim\,$24.5~mag, in the GOODS-N field. The expected SFRs of
100--200~M$_{\odot}$ yr$^{-1}$ implies that these LBG candidates at
\zdrop\ are likely to be the progenitors of massive LBGs found at
z$\,\simeq\,$5. The number density of such luminous LBGs puts strong
constraints on the bright-end of the luminosity function at \zdrop,
consistent within the theoretical limit based on the halo mass
function. Deep NIR spectroscopic observations will help to confirm
these bright candidates as well as shed light on the nature of such
luminous LBGs at \zdrop.

\acknowledgments
We thank the referee for helpful comments and suggestions that
significantly improved this paper. We acknowledge valuable comments
and suggestions from Mark Dickinson. We are grateful to Kyle
Penner and Shoubaneh Hemmati for their help in reducing the Keck
spectroscopy data. NPH acknowledges support provided by NASA through
grants HST-GO-11702.02-A and HST-GO-11359.08-A from the Space
Telescope Science Institute, which is operated by AURA, Inc., under
NASA contract NAS 5-26555.


\clearpage

\clearpage

\clearpage

\clearpage

\clearpage

\clearpage

\clearpage

\clearpage



\begin{deluxetable}{ccccccc}
\tablewidth{0pt}
\tabletypesize{\footnotesize}
\tablecaption{GOODS-N Near-IR data \label{tab:data}}
\tablenum{1}
\tablehead{\colhead{Telescope/Camera} & \colhead{Filters} &
  \colhead{Area Coverage} & \colhead{Image Pixel Scale} & Exposure &
  \ks\ Magnitude & Typical \\
  &      &   \colhead{ (\sqarc)} &      \colhead{(arcsec
    pixel$^{-1}$)} & Time$^d$ & Limit$^e$ (mag) & 
  Seeing (\arcsec)}

\startdata
Subaru/\moircs & $J$, $H$, \ks & 109                      &0.15$^a$ &
$\sim$2 hrs  & $\sim$24.0 &  0.6-0.8\\ 
CFHT/\wcam & $J$, \ks & 1040$^b$                      & 0.30   & $\sim$3 hrs &
$\sim$24.0 & 0.6-0.8 \\
Combined$^c$ & $J$, \ks & 1040$^b$                      & 0.30 & $\sim$5 hrs &
$\sim$24.5 & 0.6-0.8 
\enddata

\tablenotetext{a}{Native pixel scale is 0.117\arcsec\ pixel$^{-1}$ but to be
  compatible with other GOODS multi-wavelength data we have used
  0.15\arcsec\ pixel$^{-1}$.}
\tablenotetext{b}{Total WIRCam coverage. For the \acsz-dropout selection, we have used
  only GOODS-N ACS coverage area of $\sim$169~\sqarc.}
\tablenotetext{c}{MOIRCS and WIRCam images are combined
  together to increase the depth in the MOIRCS covered area, which is
  inside the GOODS-N ACS covered region.}
\tablenotetext{d}{80\% of the area has at least this exposure time.}
\tablenotetext{e}{50\% completeness limit (3$\sigma$)  from \figref{fig:ncounts}.}
\end{deluxetable}


\begin{deluxetable}{cc}
\tablewidth{0pt}
\tablecaption{\texttt{SExtractor} Parameters used for \ks-selected
  and $J$-selected catalogs. The values in parenthesis are used for
  $J$-selected catalog with other parameters same as \ks-selected catalog. \label{tab:param}}
\tablenum{2}
\tablehead{\colhead{Parameter} & \colhead{Value}}
\startdata
DETECT\_MINAREA 	& 4 (2) \\
DETECT\_THRESH    	& 2.0 (1.0)  \\
ANALYSIS\_THRESH  	& 2.0 (1.0)  \\
FILTER           	& Y \\
FILTER\_NAME		& gauss\_3.0\_5x5.conv \\
DEBLEND\_NTHRESH  	& 32 \\
DEBLEND\_MINCONT  	& 0.0001 \\
CLEAN            	& Y \\
CLEAN\_PARAM      	& 1.0  (0.3) \\
SEEING\_FWHM      	& 0.8 \\
BACK\_SIZE        	& 80 \\
BACK\_FILTERSIZE  	& 3 \\
BACKPHOTO\_TYPE   	& LOCAL \\
BACKPHOTO\_THICK  	& 12 \\
WEIGHT\_TYPE      	& MAP\_RMS 
\enddata
\end{deluxetable}


\begin{deluxetable}{cccccccccccccccc}
\tablewidth{0pt}
\rotate
\tabletypesize{\footnotesize}
\tablecaption{Photometry of two \acsz-dropout candidates. \label{tab:phot}}
\tablenum{3}
\tablehead{\colhead{ID} & \colhead{RA} & \colhead{DEC} & \colhead{$B_{\rm 435}^a$} & \colhead{$V_{\rm 606}^a$} & \colhead{$i_{\rm 775}^a$} &
  \colhead{$z_{\rm 850}^a$} & \colhead{$J^a$} & \colhead{$H^b$} & \colhead{$K_{\rm s}^a$} & \colhead{m$_{3.6}^a$}
  & \colhead{m$_{4.5}^a$} & \colhead{m$_{5.8}^b$} & \colhead{m$_{8.0}^b$} & Stellarity$^c$ & z$_{phot}$}

\startdata
A & 189.07061 & 62.20895 & $>$28.0 & 31.40  & $>$28.0  & 27.07 & 23.22 & 23.07 & 22.53  & 22.18  & 22.24 & 22.10 & 22.04 & 0.54 & 7.43 \\
  &           &          &   \nodata      & $\pm$3.26 & \nodata  & $\pm$0.16 & $\pm$0.06 &
   $\pm$0.17 & $\pm$0.05 & $\pm$0.01 & $\pm$0.02  & $\pm$0.20 & $\pm$0.21 \\ 
B & 189.08286 & 62.15940 & $>$28.0 & 29.90  & 28.09  & 26.03 & 23.94  &
24.29 & 23.82  & 23.62  & 24.25 & 24.19 & $>$23.75 & 0.13 & 6.92 \\
   &                   &                 & \nodata  & $\pm$0.81 & $\pm$0.31 & $\pm$0.07 & $\pm$0.09 &
   $\pm$0.54 & $\pm$0.14 & $\pm$0.03 & $\pm$0.08 & $\pm$0.55 & \nodata 
\enddata

\tablenotetext{a}{These magnitudes are based on TFIT technique
  \citep{laid07,dahl10}.}
\tablenotetext{b}{These magnitudes are based on \texttt{SExtractor} \texttt{MAG\_AUTO} estimates.}
\tablenotetext{c}{Star/Galaxy separation based on \texttt{SExtractor} CLASS\_STAR index (1=star) measured in $J$-band.}

\end{deluxetable}

\end{document}